\documentclass{article}
\usepackage{amsmath}

\input{tcilatex}
\begin{document}

\author{Nick Laskin\thanks{%
Email address: nlaskin@rocketmail.com}}
\title{\textbf{Principles of Fractional Quantum Mechanics}\\
}
\date{TopQuark Inc.\\
Toronto, ON, M6P 2P2 \\
Canada}
\maketitle

\begin{abstract}
A review of fundamentals and physical applications of fractional quantum
mechanics has been presented.

Fundamentals cover fractional Schr\"{o}dinger equation, quantum Riesz
fractional derivative, path integral approach to fractional quantum
mechanics, hermiticity of the Hamilton operator, parity conservation law and
the current density. Applications of fractional quantum mechanics cover
dynamics of a free particle, new representation for a free particle quantum
mechanical kernel, infinite potential well, bound state in $\delta $%
-potential well, linear potential, fractional Bohr atom and fractional
oscillator.

We also review fundamentals of the L\'{e}vy path integral approach to
fractional statistical mechanics.

\textit{PACS }number(s): 05.40.Fb, 05.30.-d, 03.65.Sq
\end{abstract}

\section{Introduction}

Classical mechanics and quantum mechanics are based on the assumption that
the Hamilton function has the form

\begin{equation}
H(\mathbf{p},\mathbf{r})=\frac{\mathbf{p}^2}{2m}+V(\mathbf{r}),  \label{eq1}
\end{equation}

where $\mathbf{p}$ and $\mathbf{r}$ are the momentum and space coordinate of
a particle with mass $m$, and $V(\mathbf{r})$ is the potential energy. In
quantum mechanics, $\mathbf{p}$ and $\mathbf{r}$ should be considered as
quantum mechanical operators $\widehat{\mathbf{p}}$ and $\widehat{\mathbf{r}}
$. Then the Hamiltonian function $H(\mathbf{p},\mathbf{r})$ becomes the
Hamilton operator $\widehat{H}(\widehat{\mathbf{p}},\widehat{\mathbf{r}}),$

\begin{equation}
\widehat{H}(\widehat{\mathbf{p}},\widehat{\mathbf{r}})=\frac{\widehat{%
\mathbf{p}}^2}{2m}+\widehat{V}(\widehat{\mathbf{r}}),  \label{eq2}
\end{equation}

where $\widehat{V}(\widehat{\mathbf{r}})$ is the potential energy operator$.$

The square dependence on the momentum in Eqs.(\ref{eq1}) and (\ref{eq2}) is
empirical physical fact. However an attempt to get insight on the
fundamentals behind this fact posts the question: are there other forms of
kinematic term in Eqs.(\ref{eq1}) and (\ref{eq2}) which do not contradict
the fundamental principles of classical mechanics and quantum mechanics? A
convenient theoretical physics approach to answer this question is the
Feynman path integral approach to quantum mechanics \cite{Feynman}, as it
was first observed by Laskin \cite{Laskin1}. Indeed, the Feynman path
integral is the integral over Brownian-like paths. Brownian motion is a
special case of so-called $\alpha $-stable probability distributions
developed by L\'{e}vy \cite{Levy} and Khintchine \cite{Khintchine}. In mid
1930's they posed the question: Does the sum of $N$ independent identically
distributed random quantities $X=X_{1}+X_{2}...+X_{N}$ have the same
probability distribution $p_{N}(X)$ (up to scale factor) as the individual
steps $p_{i}(X_{i})$, $i=1,...N$? The traditional answer is that each $%
p_{i}(X_{i})$ should be a Gaussian, because of the central limit theorem. In
other words, a sum of $N$ Gaussians is again a Gaussian, but with $N$ times
the variance of the original. L\'{e}vy and Khintchine proved that there
exist the possibility to generalize the central limit theorem. They
discovered a class of non-Gaussian $\alpha $-stable (stable under summation)
probability distributions. Each $\alpha $-stable distribution has a
stability index $\alpha $, often called the L\'{e}vy index $0<\alpha \leq 2$%
. When $\alpha =2$ L\'{e}vy motion is transformed into Brownian motion.

An option to develop the path integral over L\'{e}vy paths was discussed by
Kac \cite{Kac}, who pointed out that the L\'{e}vy path integral generates a
functional measure in the space of left (or right) continuos functions
(paths) having only discontinuities of the first kind. The path integral
over the L\'{e}vy paths has first been introduced and elaborated with
applications to fractional quantum mechanics and fractional statistical
mechanics by Laskin (see, \cite{Laskin1}, \cite{Laskin2}-\cite{Laskin5}). He
followed the framework of the Feynman space-time vision of quantum
mechanics, but instead of the Brownian-like quantum mechanical trajectories,
Laskin used the L\'{e}vy-like ones. If the fractal dimension (for definition
of fractal dimension, see \cite{Mandelbrot}, \cite{Feder}) of the Brownian
path is $\mathrm{d}_{\mathrm{fractal}}^{(Brownian)}=$2, then the L\'{e}vy
path has fractal dimension $\mathrm{d}_{\mathrm{fractal}}^{(L\acute{e}%
vy)}=\alpha $, where $\alpha $ is so-called the L\'{e}vy index, $1<\alpha
\leq 2$. The L\'{e}vy index $\alpha $ becomes a new fundamental parameter in
fractional quantum and classical mechanics similar to $\mathrm{d}_{\mathrm{%
fractal}}^{(Brownian)}=2$ being a fundamental parameter in standard quantum
and classical mechanics. The difference between the fractal dimensions of
the Brownian and L\'{e}vy paths leads to different physics. In fact,
fractional quantum dynamics is generated by the Hamiltonian function $%
H_{\alpha }(\mathbf{p},\mathbf{r})$ of the form \cite{Laskin1}, \cite%
{Laskin2}-\cite{Laskin5}

\begin{equation}
H_{\alpha }(\mathbf{p},\mathbf{r})=D_{\alpha }|\mathbf{p}|^{\alpha }+V(%
\mathbf{r}),\qquad 1<\alpha \leq 2,  \label{eq3}
\end{equation}

with substitutions $\mathbf{p\rightarrow }\widehat{\mathbf{p}}$, $\mathbf{%
r\rightarrow }$ $\widehat{\mathbf{r}}$, and $D_{\alpha }$ is the generalized
coefficient, the physical dimension of which is $[D_{\alpha }]=\mathrm{erg}%
^{1-\alpha }\cdot \mathrm{cm}^{\alpha }\cdot \mathrm{sec}^{-\alpha }.$ One
can say that Eq.(\ref{eq3}) is a natural generalization of the well-known
Eq.(\ref{eq1}). When $\alpha =2$, $D_{\alpha }=1/2m$ and Eq.(\ref{eq3}) is
transformed into Eq.(\ref{eq1}) \cite{Laskin1}. As a result, the fractional
quantum mechanics based on the L\'{e}vy path integral generalizes the
standard quantum mechanics based on the well-known Feynman path integral.
Indeed, if the path integral over Brownian trajectories leads to the
well-known Schr\"{o}dinger equation, then the path integral over L\'{e}vy
trajectories leads to the fractional Schr\"{o}dinger equation. The
fractional Schr\"{o}dinger equation is a new fundamental equation of quantum
physics and it includes the space derivative of order $\alpha $ instead of
the second $(\alpha =2)$ order space derivative in the standard Schr\"{o}%
dinger equation. Thus, the fractional Schr\"{o}dinger equation is the
fractional differential equation in accordance with modern terminology (see,
for example, \cite{Oldham}-\cite{Podlubny}). This is the main point for the
term \textit{fractional Schr\"{o}dinger equation} or for more general term 
\textit{fractional quantum mechanics}, FQM \cite{Laskin1}, \cite{Laskin2}.
When L\'{e}vy index $\alpha =2$,\ L\'{e}vy motion becomes Brownian motion.
Thus, FQM includes standard QM as a particular Gaussian case at $\alpha =2$.
The quantum mechanical path integral over the L\'{e}vy paths \cite{Laskin1}
at $\alpha =2$ becomes the Feynman path integral \cite{Feynman}.

In the limit case $\alpha =2$ the fundamental equations of fractional
quantum mechanics are transformed into the well-known equations of standard
quantum mechanics \cite{Feynman}, \cite{Landau}, \cite{Kleinert}.

\section{Fractional Schr\"{o}dinger equation}

\subsection{Quantum Riesz fractional derivative}

Equation (\ref{eq1}) lets us conclude that the energy $E$ of a particle of
mass $m$ under the influence of the potential $V(\mathbf{r})$ is given by

\begin{equation}
E=\frac{\mathbf{p}^{2}}{2m}+V(\mathbf{r}).  \label{eq4}
\end{equation}

To obtain the Schr\"{o}dinger equation we introduce the operators following
the well-known procedure,

\begin{equation}
E\rightarrow i\hbar \frac \partial {\partial t},\qquad \qquad \mathbf{p}%
\rightarrow -i\hbar \mathbf{\nabla },  \label{eq5}
\end{equation}

where $\mathbf{\nabla }=\partial /\partial \mathbf{r}$ and $\hbar $ is the
Planck's constant. Further, substituting transformation (\ref{eq5}) into Eq.(%
\ref{eq1}) and applying them to the wave function $\psi (\mathbf{r},t)$
yields

\begin{equation}
i\hbar \frac{\partial \psi (\mathbf{r},t)}{\partial t}=-\frac{\hbar ^2}{2m}%
\Delta \psi (\mathbf{r},t)+V(\mathbf{r})\psi (\mathbf{r},t),  \label{eq6}
\end{equation}

here $\Delta =\mathbf{\nabla }\cdot \mathbf{\nabla }$ is the Laplacian.
Thus, we obtain the Schr\"{o}dinger equation \cite{Landau}.

By repeating the same consideration to Eq.(\ref{eq3}) we find the fractional
Schr\"{o}dinger equation \cite{Laskin1}, \cite{Laskin4}

\begin{equation}
i\hbar \frac{\partial \psi (\mathbf{r},t)}{\partial t}=D_\alpha (-\hbar
^2\Delta )^{\alpha /2}\psi (\mathbf{r},t)+V(\mathbf{r})\psi (\mathbf{r}%
,t),\qquad 1<\alpha \leq 2,  \label{eq7}
\end{equation}

with 3D generalization of the fractional quantum Riesz derivative $(-\hbar
^{2}\Delta )^{\alpha /2}$ introduced by

\begin{equation}
(-\hbar ^2\Delta )^{\alpha /2}\psi (\mathbf{r},t)=\frac 1{(2\pi \hbar
)^3}\int d^3pe^{i\frac{\mathbf{pr}}\hbar }|\mathbf{p}|^\alpha \varphi (%
\mathbf{p},t),  \label{eq8}
\end{equation}

where the wave functions in space $\psi (\mathbf{r},t)$ and momentum $%
\varphi (\mathbf{p},t)$ representations are related each to other by the 3D
Fourier transforms

\begin{equation}
\psi (\mathbf{r},t)=\frac{1}{(2\pi \hbar )^{3}}\int d^{3}pe^{i\frac{\mathbf{%
pr}}{\hbar }}\varphi (\mathbf{p},t),\qquad \varphi (\mathbf{p},t)=\int
d^{3}re^{-i\frac{\mathbf{pr}}{\hbar }}\psi (\mathbf{r},t).  \label{eq9}
\end{equation}

The 3D fractional Schr\"{o}dinger equation Eq.(\ref{eq7}) has the following
operator form

\begin{equation*}
i\hbar \frac{\partial \psi (\mathbf{r},t)}{\partial t}=\widehat{H}_\alpha (%
\widehat{\mathbf{p}},\widehat{\mathbf{r}})\psi (\mathbf{r},t),
\end{equation*}

where fractional Hamilton operator $\widehat{H}_{\alpha }(\widehat{\mathbf{p}%
},\widehat{\mathbf{r}})$ results from Eq.(\ref{eq3}) with quantum-mechanical
operators $\widehat{\mathbf{p}}$ and $\widehat{\mathbf{r}}$ substituted
instead of $\mathbf{p}$ and $\mathbf{r}$,

\begin{equation}
\widehat{H}_{\alpha }(\widehat{\mathbf{p}},\widehat{\mathbf{r}})=D_{\alpha }|%
\widehat{\mathbf{p}}|^{\alpha }+V(\widehat{\mathbf{r}}),\qquad 1<\alpha \leq
2.  \label{eq10a}
\end{equation}

The 1D fractional Schr\"{o}dinger equation has the form \cite{Laskin1}, \cite%
{Laskin2}-\cite{Laskin4}

\begin{equation}
i\hbar \frac{\partial \psi (x,t)}{\partial t}=-D_\alpha (\hbar \nabla
)^\alpha \psi (x,t)+V(x)\psi (x,t),\qquad 1<\alpha \leq 2,  \label{eq10}
\end{equation}

where ($\hbar \nabla )^{\alpha }$ is the quantum Riesz fractional derivative%
\footnote{%
The Riesz fractional derivative was originally introduced in \cite{Riesz}}.
The quantum Riesz fractional derivative is defined by the following way \cite%
{Laskin1}, \cite{Laskin2}

\begin{equation}
(\hbar \nabla )^\alpha \psi (x,t)=-\frac 1{2\pi \hbar }\int\limits_{-\infty
}^\infty dpe^{i\frac{px}\hbar }|p|^\alpha \varphi (p,t),  \label{eq11}
\end{equation}

where $\varphi (p,t)$ is the Fourier transform of the wave function $\psi
(x,t)$ given by

\begin{equation}
\varphi (p,t)=\int\limits_{-\infty }^\infty dxe^{-i\frac{px}\hbar }\psi
(x,t),  \label{eq12}
\end{equation}

and reciprocally

\begin{equation*}
\psi (x,t)=\frac 1{2\pi \hbar }\int\limits_{-\infty }^\infty dpe^{i\frac{px}%
\hbar }\varphi (p,t).
\end{equation*}

It is easy to see that Eq.(\ref{eq10}) can be rewritten in the operator
form, namely

\begin{equation}
i\hbar \frac{\partial \psi }{\partial t}=H_\alpha \psi ,  \label{eq13}
\end{equation}

where $H_\alpha $ is the fractional Hamilton operator

\begin{equation}
H_\alpha =-D_\alpha (\hbar \nabla )^\alpha +V(x).  \label{eq14}
\end{equation}

For the special case when $\alpha =2$ and $D_{2}=1/2m$ (see, for details 
\cite{Laskin1}, \cite{Laskin2}), where $m$ is the particle mass, Eqs.(\ref%
{eq7}) and (\ref{eq10}) are transformed into the well-known 3D and 1D Schr%
\"{o}dinger equations \cite{Landau}.

\subsection{The hermiticity of the fractional Hamilton operator}

The fractional Hamiltonian $H_\alpha $ given by Eq.(\ref{eq14}) is the
Hermitian operator in the space with scalar product

\begin{equation}
(\phi ,\chi )=\int\limits_{-\infty }^\infty dx\phi ^{*}(x,t)\chi (x,t).
\label{eq15}
\end{equation}

To prove the hermiticity of $H_{\alpha }$ let us note that in accordance
with the definition of the quantum Riesz fractional derivative given by Eq.(%
\ref{eq11}) there exists the integration-by parts formula

\begin{equation}
(\phi ,(\hbar \nabla )^\alpha \chi )=((\hbar \nabla )^\alpha \phi ,\chi ).
\label{eq16}
\end{equation}

The average energy of a fractional quantum system with Hamiltonian $%
H_{\alpha }$ is

\begin{equation}
E_\alpha =\int\limits_{-\infty }^\infty dx\psi ^{*}(x,t)H_\alpha \psi (x,t).
\label{eq17}
\end{equation}

Taking into account Eq.(\ref{eq16}) we have

\begin{equation*}
E_\alpha =\int\limits_{-\infty }^\infty dx\psi ^{*}(x,t)H_\alpha \psi
(x,t)=\int\limits_{-\infty }^\infty dx(H_\alpha ^{+}\psi (x,t))^{*}\psi
(x,t)=E_\alpha ^{*}.
\end{equation*}

As a physical consequence, the energy of the system is real. Thus, the
fractional Hamiltonian $H_{\alpha }$ defined by Eq.(\ref{eq14}) is the
Hermitian or self-adjoint operator in the space with the scalar product
defined by Eq.(\ref{eq15}) \cite{Laskin3}, \cite{Laskin4}

\begin{equation}
(H_\alpha ^{+}\phi ,\chi )=(\phi ,H_\alpha \chi ).  \label{eq18}
\end{equation}

The generalization of the proof of hermiticity for 3D case is
straightforward. Note that Eq.(\ref{eq10}) leads to the important equation

\begin{equation}
\frac \partial {\partial t}\int dx\psi ^{*}(x,t)\psi (x,t)=0,  \label{eq19}
\end{equation}

which shows that the wave function remains normalized, if it is normalized
once. Multiplying Eq.(\ref{eq10}) from the left by $\psi ^{\ast }(x,t)$ and
the conjugate complex of Eq.(\ref{eq10}) by $\psi (x,t)$ and then
subtracting the two resultant equations finally yield

\begin{equation*}
i\hbar \frac \partial {\partial t}\left( \psi ^{*}(x,t)\psi (x,t)\right)
=\psi ^{*}(x,t)H_\alpha \psi (x,t)-\psi (x,t)H_\alpha ^{*}\psi ^{*}(x,t).
\end{equation*}

Integrating this relation over all values of the space variable and using
the fact that the operator $H_{\alpha }$ is self-adjoint, we find Eq.(\ref%
{eq19}). The above consideration can be easily generalized to 3D case.

\subsection{The parity conservation law for the fractional quantum mechanics}

It follows from the definition (\ref{eq11}) of the quantum Riesz fractional
derivative that

\begin{equation}
(\hbar \nabla )^{\alpha }\exp \{i\frac{px}{\hbar }\}=-|p|^{\alpha }\exp \{i%
\frac{px}{\hbar }\},  \label{eq20p}
\end{equation}

in other words, the function $\exp \{ipx/\hbar \}$ is the eigenfunction of
the quantum Riesz fractional operator $(\hbar \nabla )^\alpha $ with
eigenvalue $-|p|^\alpha $.

The 3D generalization is straightforward,

\begin{equation}
(-\hbar ^{2}\Delta )^{\alpha /2}\exp \{i\frac{\mathbf{px}}{\hbar }\}=|%
\mathbf{p}|^{\alpha }\exp \{i\frac{\mathbf{px}}{\hbar }\},  \label{eq21p}
\end{equation}

the function $\exp \{i\mathbf{px}/\hbar \}$ is the eigenfunction of 3D
quantum Riesz fractional operator $(-\hbar ^{2}\Delta )^{\alpha /2}$ with
eigenvalue $|\mathbf{p}|^{\alpha }$.

Thus, the operators $(\hbar \nabla )^\alpha $ and $(-\hbar ^2\Delta
)^{\alpha /2}$ are the symmetrized fractional derivative, that is

\begin{equation}
(\hbar \nabla _{x})^{\alpha }...=(\hbar \nabla _{-x})^{\alpha }...,
\label{eq22p}
\end{equation}

\begin{equation}
(-\hbar ^{2}\Delta _{\mathbf{r}})^{\alpha /2}...=(-\hbar ^{2}\Delta _{-%
\mathbf{r}})^{\alpha /2}....  \label{eq23p}
\end{equation}

Because of the properties (\ref{eq16}) and (\ref{eq17}) the fractional
Hamiltonian $H_\alpha $ (see, for example Eqs.(\ref{eq5}) or (\ref{eq8}))
remains invariant under \textit{inversion} transformation. Inversion, or to
be precise, spatial inversion consists in the simultaneous change in sign of
all three spatial coordinates

\begin{equation}
\mathbf{r}\rightarrow -\mathbf{r},\qquad x\rightarrow -x,\quad y\rightarrow
-y,\quad z\rightarrow -z.  \label{eq18p}
\end{equation}

Let us denote the inversion operator by $\widehat{P}$. The inverse symmetry
is the fact that $\widehat{P}$ and the fractional Hamiltonian $H_\alpha $
commute,

\begin{equation}
\widehat{P}H_{\alpha }=H_{\alpha }\widehat{P}.  \label{eq19p}
\end{equation}

We can divide the wave functions of quantum mechanical states with a
well-defined eigenvalue of the operator $\widehat{P}$ into two classes; (i)
functions which are not changed when acted upon by the inversion operator, $%
\widehat{P}\psi _{+}(\mathbf{r})=\psi _{+}(\mathbf{r})$ the corresponding
states are called even states; (ii) functions which change sign under the
action of the inversion operator, $\widehat{P}\psi _{-}(\mathbf{r})=-\psi
_{-}(\mathbf{r})$ the corresponding states are called odd states. Eq.(\ref%
{eq19p}) express the \textquotedblright parity conservation
law\textquotedblright\ for the FQM \cite{Laskin4}; if the state of a closed
fractional quantum mechanical system has a given parity (i.e. if it is even,
or odd), then this parity is conserved.

\subsection{The current density}

By multiplying Eq.(\ref{eq7}) from left by $\psi ^{\ast }(\mathbf{r},t)$ and
the conjugate complex of Eq.(\ref{eq7}) by $\psi (\mathbf{r},t)$ and
subtracting the two resultant equations yield

\begin{equation}
\frac \partial {\partial t}\int d^3r\left( \psi ^{*}(\mathbf{r},t)\psi (%
\mathbf{r},t)\right) =  \label{eq25}
\end{equation}

\begin{equation*}
\frac{D_\alpha }{i\hbar }\int d^3r\left( \psi ^{*}(\mathbf{r},t)(-\hbar
^2\Delta )^{\alpha /2}\psi (\mathbf{r},t)-\psi (\mathbf{r},t)(-\hbar
^2\Delta )^{\alpha /2}\psi ^{*}(\mathbf{r},t)\right) .
\end{equation*}

From this integral relationship we are led to the following well-known
differential equation

\begin{equation}
\frac{\partial \rho (\mathbf{r},t)}{\partial t}+\mathrm{div}\mathbf{j}(%
\mathbf{r},t)=0,  \label{eq26}
\end{equation}

where $\rho (\mathbf{r},t)=\psi ^{\ast }(\mathbf{r},t)\psi (\mathbf{r},t)$
is the quantum mechanical probability density and the vector $\mathbf{j}(%
\mathbf{r},t)$ can be called by the fractional probability current density
vector

\begin{equation}
\mathbf{j}(\mathbf{r},t)=\frac{D_\alpha \hbar }i\left( \psi ^{*}(\mathbf{r}%
,t)(-\hbar ^2\Delta )^{\alpha /2-1}\mathbf{\nabla }\psi (\mathbf{r},t)-\psi (%
\mathbf{r},t)(-\hbar ^2\Delta )^{\alpha /2-1}\mathbf{\nabla }\psi ^{*}(%
\mathbf{r},t)\right) ,  \label{eq27}
\end{equation}

where we use the following notation $\mathbf{\nabla =\partial /\partial r}$.
Introducing the momentum operator $\widehat{\mathbf{p}}=\frac{\hbar }{i}%
\frac{\partial }{\partial \mathbf{r}}$ we can write the vector $\mathbf{j}$
in the form \cite{Laskin4}

\begin{equation}
\mathbf{j=}D_\alpha \left( \psi (\widehat{\mathbf{p}}^2)^{\alpha /2-1}%
\widehat{\mathbf{p}}\psi ^{*}+\psi ^{*}(\widehat{\mathbf{p}}^{*2})^{\alpha
/2-1}\widehat{\mathbf{p}}^{*}\psi \right)  \label{eq28}
\end{equation}

The new fundamental Eqs.(\ref{eq27}) and (\ref{eq28}) are the fractional
generalization of the well-known equations for probability current density
vector of standard quantum mechanics \cite{Landau}.

To this end, we express Eq.(\ref{eq28}) in the terms of the velocity
operator, which is defined as follows $\widehat{\mathbf{v}}=d\widehat{%
\mathbf{r}}/dt,$ where $\widehat{\mathbf{r}}$ is the operator of coordinate.
Using the general quantum mechanical rule for differentiation of operator

\begin{equation*}
\frac d{dt}\widehat{\mathbf{r}}=\frac i\hbar [H_\alpha ,\mathbf{r}],
\end{equation*}

we have

\begin{equation*}
\widehat{\mathbf{v}}=\frac{i}{\hbar }(H_{\alpha }\mathbf{r-r}H_{\alpha }),
\end{equation*}

Further, with help of the equation $\mathrm{f}(\widehat{\mathbf{p}})\mathbf{r%
}-\mathbf{r}\mathrm{f}(\widehat{\mathbf{p}})=-i\hbar \partial \mathrm{%
f/\partial }\mathbf{p}$, which holds for any function $\mathrm{f}(\widehat{%
\mathbf{p}})$ of the momentum operator, and taking into account Eq.(\ref%
{eq10a}) for the Hamiltonian operator $\widehat{H}_{\alpha }(\widehat{%
\mathbf{p}},\widehat{\mathbf{r}})$ we obtain the equation for the velocity
operator

\begin{equation}
\widehat{\mathbf{v}}=\alpha D_{\alpha }|\widehat{\mathbf{p}}^{2}|^{\alpha
/2-1}\widehat{\mathbf{p}},  \label{eq30c}
\end{equation}

here $\widehat{\mathbf{p}}$ is the momentum operator. By comparing of Eqs.(%
\ref{eq28}) and (\ref{eq30c}) we finally conclude that

\begin{equation}
\mathbf{j=}\frac{1}{\alpha }\left( \psi \widehat{\mathbf{v}}\psi ^{\ast
}+\psi ^{\ast }\widehat{\mathbf{v}}\psi \right) ,\qquad 1<\alpha \leq 2.
\label{eq31c}
\end{equation}

To get the probability current density equal to $1$ (the current when one
particle passes through unit area per unit time) the wave function of a free
particle has to be normalized as

\begin{equation}
\psi (\mathbf{r},t)=\sqrt{\frac{\alpha }{2\mathrm{v}}}\exp \{\frac{i}{\hbar }%
\mathbf{pr}-\frac{i}{\hbar }Et\},\qquad E=D_{\alpha }|\mathbf{p}|^{\alpha
},\qquad 1<\alpha \leq 2,  \label{eq32c}
\end{equation}

where $\mathrm{v}$ is the particle velocity, $\mathrm{v}=\alpha D_{\alpha
}p^{\alpha -1}$. Then we have

\begin{equation}
\mathbf{j=}\frac{\mathbf{v}}{\mathrm{v}},\qquad \mathbf{v}=\alpha D_{\alpha
}|\mathbf{p}^{2}|^{\frac{\alpha }{2}-1}\mathbf{p,}  \label{eq33c}
\end{equation}

that is, the vector $\mathbf{j}$ is indeed the unit vector.

Equations (\ref{eq27})-(\ref{eq33c}) are the fractional generalization of
the well-known equations for probability current density vector and velocity
vector of the standard quantum mechanics \cite{Landau}.

\subsection{The time-independent fractional Schr\"{o}dinger equation}

The special case when the Hamiltonian $H_{\alpha }$ does not depend
explicitly on the time is of great importance for physical applications. It
is easy to see that in this case there exist the special solution of the
fractional Schr\"{o}dinger equation (\ref{eq10}) of the form

\begin{equation}
\psi (x,t)=e^{-(i/\hbar )Et}\phi (x),  \label{eq20}
\end{equation}

where $\phi (x)$ satisfies

\begin{equation}
H_\alpha \phi (x)=E\phi (x),  \label{eq21}
\end{equation}

or

\begin{equation}
-D_\alpha (\hbar \nabla )^\alpha \phi (x)+V(x)\phi (x)=E\phi (x),\qquad
1<\alpha \leq 2.  \label{eq22}
\end{equation}

The equation (\ref{eq22}) we call by the time-independent (or stationary)
fractional Schr\"{o}dinger equation \cite{Laskin3}, \cite{Laskin4}. We see
from Eq.(\ref{eq20}) that the wave function $\psi (x,t)$ oscillates with a
definite frequency. The frequency with which a wave function oscillates
corresponds to the energy. Therefore, we say that when the fractional wave
function $\psi (x,t)$ is of this special form, the state has a definite
energy $E$. The probability to find a particle at $x$ is the absolute square
of the wave function \TEXTsymbol{\vert}$\psi $\TEXTsymbol{\vert}$^{2}$. In
view of Eq.(\ref{eq20}) this is equal to \TEXTsymbol{\vert}$\phi $%
\TEXTsymbol{\vert}$^{2}$and does not depend upon the time. That is, the
probability of finding the particle in any location is independent of the
time. We say under these circumstances that the system is in a stationary
state - stationary in the sense that there is no variation in the
probabilities as a function of time.

\section{Path integral}

If a particle at an initial time $t_{a}$ starts from the point $x_{a}$ and
goes to a final point $x_{b}$ at time $t_{b}$, we will say simply that the
particle goes from $a$ to $b$ and its trajectory (path)\footnote{%
For simplicity, here we consider one dimensional motion.} $x(t)$ will have
the property that $x(t_{a})=x_{a}$ and $x(t_{b})=x_{b}$. In quantum
mechanics, then, we will have an quantum-mechanical amplitude, often called
a kernel, which we may write $K(x_{b}t_{b}|x_{a}t_{a})$, to get from the
point $a$ to the point $b$. This will be the sum over all of the
trajectories that go between that end points and of a contribution from each 
\cite{Feynman}. For the one dimensional version of the Hamiltonian Eq.(\ref%
{eq3})%
\begin{equation}
H_{\alpha }(p,x)=D_{\alpha }|p|^{\alpha }+V(x,t),  \label{eq23pi}
\end{equation}%
following consideration provided in \cite{Laskin1}, \cite{Laskin2}, we come
to the definition of the kernel $K(x_{b}t_{b}|x_{a}t_{a})$ in terms of path
integral in the phase space representation

\begin{equation}
K(x_{b}t_{b}|x_{a}t_{a})=\underset{N\rightarrow \infty }{\lim }%
\int\limits_{-\infty }^{\infty }dx_{1}...dx_{N-1}\frac{1}{(2\pi \hbar )^{N}}%
\int\limits_{-\infty }^{\infty }dp_{1}...dp_{N}\times  \label{eq26pi}
\end{equation}

\begin{equation*}
\exp \left\{ \frac{i}{\hbar }\sum\limits_{j=1}^{N}p_{j}(x_{j}-x_{j-1})\right%
\} \times \exp \left\{ -\frac{i}{\hbar }D_{\alpha }\varepsilon
\sum\limits_{j=1}^{N}|p_{j}|^{\alpha }-\frac{i}{\hbar }\varepsilon
\sum\limits_{j=1}^{N}V(x_{j},j\varepsilon )\right\} ,
\end{equation*}

here $\varepsilon =(t_{b}-t_{a})/N$ , $x_{j}=x(j\varepsilon ),$ $%
p_{j}=p(j\varepsilon )$ and $x_{0}=x_{a}$, $x_{N}=x_{b}$. Then in the
continuum limit $N\rightarrow \infty ,\quad \varepsilon \rightarrow 0$ we
have

\begin{equation}
K_L(x_bt_b|x_at_a)=  \label{eq42}
\end{equation}

\begin{equation*}
\int\limits_{x(t_{a})=x_{a}}^{x(t_{b})=x_{b}}\mathrm{D}x(\tau
)\int\limits_{{}}^{{}}\mathrm{D}p(\tau )\exp \left\{ \frac{i}{\hbar }%
\int\limits_{t_{a}}^{t_{b}}d\tau \lbrack p(\tau )\overset{\cdot }{x}(\tau
)-H_{\alpha }(p(\tau ),x(\tau ),\tau )]\right\} ,
\end{equation*}

where $\overset{\cdot }{x}$ denotes the time derivative, $H_{\alpha }(p(\tau
),x(\tau ),\tau )$ is the fractional Hamiltonian given by Eq.(\ref{eq23pi})
with the replacement $p\rightarrow p(\tau )$, $x\rightarrow x(\tau )$ and $%
\{p(\tau ),x(\tau )\}$ is the particle trajectory in phase space, and,
finally, the phase space path integral $\int%
\limits_{x(t_{a})=x_{a}}^{x(t_{b})=x_{b}}\mathrm{D}x(\tau
)\int\limits_{{}}^{{}}\mathrm{D}p(\tau )...$ is given by

\begin{equation}
\int\limits_{x(t_{a})=x_{a}}^{x(t_{b})=x_{b}}\mathrm{D}x(\tau
)\int\limits_{{}}^{{}}\mathrm{D}p(\tau )...=  \label{eq41pi}
\end{equation}

\begin{equation*}
=\underset{N\rightarrow \infty }{\lim }\int\limits_{-\infty }^\infty
dx_1...dx_{N-1}\frac 1{(2\pi \hbar )^N}\int\limits_{-\infty }^\infty
dp_1...dp_N\times
\end{equation*}

\begin{equation*}
\exp \left\{ i\frac{p_{1}(x_{1}-x_{a})}{\hbar }-i\frac{D_{\alpha
}\varepsilon |p_{1}|^{\alpha }}{\hbar }\right\} \times ...\times \exp
\left\{ i\frac{p_{N}(x_{b}-x_{N-1})}{\hbar }-i\frac{D_{\alpha }\varepsilon
|p_{N}|^{\alpha }}{\hbar }\right\} ...,
\end{equation*}

The exponential in Eq.(\ref{eq42}) can be written as $\exp \{\frac{i}{\hbar }%
S_{\alpha }(p,x)\}$ if we introduce the fractional canonical classical
mechanical action $S_{\alpha }(p,x)$ for the trajectory $p(t)$, $x(t)$ in
phase space \cite{Laskin2}

\begin{equation}
S_\alpha (p,x)=\int\limits_{t_a}^{t_b}d\tau (p(\tau )\overset{\cdot }{x}%
(\tau )-H_\alpha (p(\tau ),x(\tau ),\tau ).  \label{eq44}
\end{equation}

Since the coordinates $x_{0}$, $x_{N}$ in the definition (\ref{eq41pi}) are
fixed at their initial and final points, $x_{0}=x_{a}$ and $x_{N}=x_{b}$,
the all possible trajectories in Eq.(\ref{eq42}) satisfy the boundary
condition $x(t_{b})=x_{b}$, $x(t_{a})=x_{a}$. We see that the definition
given by Eq.(\ref{eq41pi}) includes one more $p_{j}$-integrals than $x_{j}$%
-integrals. Indeed, while $x_{0}$ and $x_{N}$ are held fixed and the $x_{j}$%
-integrals are done for $j=1,...,N-1$, each increment $x_{j}-x_{j-1}$ is
accompanied by one $p_{j}$-integral for $j=1,...,N$. The above observed
asymmetry is a consequence of the particular boundary condition. Namely, the
end points are fixed in the position (coordinate) space. There exist the
possibility of proceeding in a conjugate way keeping the initial $p_{a}$ and
final $p_{b}$ momenta and fixed. The associated kernel can be derived going
through the same steps as before but working in the momentum representation
(see, for example, \cite{Kleinert}).

The kernel $K(x_{b}t_{b}|x_{a}t_{a})$ introduced by Eq.(\ref{eq42})\
describes the evolution of the quantum mechanical system

\begin{equation}
\psi (x_{b},t_{b})=\int\limits_{-\infty }^{\infty
}dx_{a}K(x_{b}t_{b}|x_{a}t_{a})\psi (x_{a},t_{a}),  \label{eq45e}
\end{equation}

where $\psi (x_{a},t_{a})$ is the wave function of the initial state (at $%
t=t_{a}$) and $\psi (x_{b},t_{b})$ is the wave function of the final state
(at $t=t_{b}$).

\subsection{Free particle}

For a free particle when $V(x,t)$, we have $H_{\alpha }(p)=D_{\alpha
}|p|^{\alpha }$ and it's easy to see that Eq.(\ref{eq26pi}) results in \cite%
{Laskin1}

\begin{equation}
K^{(0)}(x_{b}t_{b}|x_{a}t_{a})=\frac{1}{2\pi \hbar }\int\limits_{-\infty
}^{\infty }dp\cdot \exp \left\{ i\frac{p(x_{b}-x_{a})}{\hbar }-i\frac{%
D_{\alpha }|p|^{\alpha }(t_{b}-t_{a})}{\hbar }\right\} .  \label{eq38}
\end{equation}

here $K^{(0)}(x_{b}t_{b}|x_{a}t_{a})$ stands for a free particle kernel.

Taking into account Eq.(\ref{eq38}) it is easily to check on directly the
consistency condition

\begin{equation*}
K^{(0)}(x_{b}t_{b}|x_{a}t_{a})=\int\limits_{-\infty }^{\infty }dx^{\prime
}K^{(0)}(x_{b}t_{b}|x^{\prime }t^{\prime })\cdot K^{(0)}(x^{\prime
}t^{\prime }|x_{a}t_{a}).
\end{equation*}

This is a special case of the general quantum-mechanical rule: for events
occurring in succession in time the amplitudes are multiplied

\begin{equation}
K(x_{b}t_{b}|x_{a}t_{a})=\int\limits_{-\infty }^{\infty }dx^{\prime
}K(x_{b}t_{b}|x^{\prime }t^{\prime })\cdot K(x^{\prime }t^{\prime
}|x_{a}t_{a}).  \label{eq38cs}
\end{equation}

\subsubsection{Fox $H$-function representation for a free particle kernel}

Let's show how a free particle fractional quantum mechanical kernel $%
K^{(0)}(x_{b}t_{b}|x_{a}t_{a})$ defined by Eq.(\ref{eq38}) can be expressed
in the terms of the Fox $H$-function \cite{Fox}, \cite{Srivastava}, \cite%
{Mathai}. Follow by \cite{Laskin5}, we obtain the Mellin transform of the
quantum mechanical fractional kernel defined by Eq.(\ref{eq38}). Comparison
of the inverse Mellin transform with the definition of the Fox function
yields the desired expression in terms of \textquotedblright
known\textquotedblright\ function, i.e. Fox $H$-function. Note that $H$%
-function bears the name of its discoverer Fox \cite{Fox} although it have
been known at least since 1888, according to \cite{Srivastava}.

Introducing for simplicity the notations $x\equiv x_{b}-x_{a},$ $\tau \equiv
t_{b}-t_{a},$ we rewrite Eq.(\ref{eq38})

\begin{equation}
K^{(0)}(x,\tau )=\frac{1}{2\pi \hbar }\int\limits_{-\infty }^{\infty
}dp\cdot \exp \left\{ i\frac{px}{\hbar }-i\frac{D_{\alpha }|p|^{\alpha }\tau 
}{\hbar }\right\} .  \label{eq47}
\end{equation}

It is easy to see that the relation $K^{(0)}(x,\tau )=K^{(0)}(-x,\tau )$
holds. Hence, it is sufficient to consider $K_{L}^{(0)}(x,\tau )$ for $x\geq
0$ only. Further, we will use the following definitions of the Mellin
transform

\begin{equation}
\overset{\wedge }{K^{(0)}}(s,\tau )=\int\limits_{0}^{\infty
}dxx^{s-1}K^{(0)}(x,\tau ),  \label{eq48}
\end{equation}

and inverse Mellin transform

\begin{equation}
K^{(0)}(x,\tau )=\frac{1}{2\pi i}\int\limits_{c-i\infty }^{c+i\infty
}dsx^{-s}\overset{\wedge }{K^{(0)}}(s,\tau ),  \label{eq49}
\end{equation}

where the integration path is the straight line from $c-i\infty $ to $%
c+i\infty $ with $0<c<1$.

The Mellin transform of the $K^{(0)}(x,\tau )$ defined in accordance with
Eq.(\ref{eq47}) is

\begin{equation*}
\overset{\wedge }{K^{(0)}}(s,\tau )=\frac{1}{2\pi \hbar }\int\limits_{0}^{%
\infty }dx\,x^{s-1}\int\limits_{-\infty }^{\infty }dp\cdot \exp \left\{ i%
\frac{px}{\hbar }-i\frac{D_{\alpha }|p|^{\alpha }\tau }{\hbar }\right\} .
\end{equation*}

By changing of the variables of integration $p\rightarrow \left( \frac{\hbar 
}{iD_{\alpha }\tau }\right) ^{1/\alpha }\varsigma ,$ $x\rightarrow \left( 
\frac{\hbar }{iD_{\alpha }\tau }\right) ^{1/\alpha }\xi ,$ one obtains the
integrals in the complex $\varsigma $ and $\xi $ planes. Considering the
paths of integration in the $\varsigma $ and $\xi $ planes it is easy to
represent $\overset{\wedge }{K^{(0)}}(s,\tau )$ as follows

\begin{equation*}
\overset{\wedge }{K^{(0)}}(s,\tau )=
\end{equation*}

\begin{equation}
\frac 1{2\pi }\left( \frac \hbar {(\hbar /iD_\alpha \tau )^{1/\alpha
}}\right) ^{s-1}\int\limits_0^\infty d\xi \xi ^{s-1}\int\limits_{-\infty
}^\infty d\varsigma \exp \{i\varsigma \xi -|\varsigma |^\alpha \}.
\label{eq50}
\end{equation}

The integrals over $d\xi $ and $d\varsigma $ can be evaluated by using the
equation

\begin{equation}
\int\limits_0^\infty d\xi \xi ^{s-1}\int\limits_0^\infty d\varsigma \exp
\{i\varsigma \xi -\varsigma ^\alpha \}=\frac 4{s-1}\sin \frac{\pi (s-1)}%
2\Gamma (s)\Gamma (1-\frac{s-1}\alpha ),  \label{eq51}
\end{equation}

where $s-1<\alpha \leq 2$ and $\Gamma (s)$ is the gamma function\footnote{%
The gamma function $\Gamma (s)$ has familiar integral representation $\Gamma
(s)=\int\limits_{0}^{\infty }dtt^{s-1}e^{-t}$, $\mathrm{Re}s>0$.}.

Inserting Eq.(\ref{eq51}) into Eq.(\ref{eq50}) and using the functional
relations for the gamma function, $\Gamma (1-z)=-z\Gamma (-z)$ and $\Gamma
(z)\Gamma (1-z)=\pi /\sin \pi z$, yield

\begin{equation*}
\overset{\wedge }{K^{(0)}}(s,\tau )=\frac{1}{\alpha }\left( \frac{\hbar }{%
(\hbar /iD_{\alpha }\tau )^{1/\alpha }}\right) ^{s-1}\frac{\Gamma (s)\Gamma (%
\frac{1-s}{\alpha })}{\Gamma (\frac{1-s}{2})\Gamma (\frac{1+s}{2})}.
\end{equation*}

The inverse Mellin transform gives a free particle quantum mechanical kernel 
$K^{(0)}(x,\tau )$

\begin{equation*}
K^{(0)}(x,\tau )=\frac{1}{2\pi i}\int\limits_{c-i\infty }^{c+i\infty
}dsx^{-s}\overset{\wedge }{K_{L}^{(0)}}(s,\tau )=
\end{equation*}

\begin{equation*}
\frac 1{2\pi i}\frac 1\alpha \cdot \int\limits_{c-i\infty }^{c+i\infty
}ds\left( \frac \hbar {(\hbar /iD_\alpha \tau )^{1/\alpha }}\right)
^{s-1}x^{-s}\frac{\Gamma (s)\Gamma (\frac{1-s}\alpha )}{\Gamma (\frac{1-s}%
2)\Gamma (\frac{1+s}2)},
\end{equation*}

where the integration path is the straight line from $c-i\infty $ to $%
c+i\infty $ with $0<c<1$. Replacing $s$ by $-s$ we obtain

\begin{equation}
K^{(0)}(x,\tau )=  \label{eq52}
\end{equation}

\begin{equation*}
\frac{1}{\alpha }\left( \frac{\hbar }{(\hbar /iD_{\alpha }\tau )^{1/\alpha }}%
\right) ^{-1}\frac{1}{2\pi i}\int\limits_{-c-i\infty }^{-c+i\infty }ds\left( 
\frac{1}{\hbar }\left( \frac{\hbar }{iD_{\alpha }\tau }\right) ^{1/\alpha
}x\right) ^{s}\frac{\Gamma (-s)\Gamma (\frac{1+s}{\alpha })}{\Gamma (\frac{%
1+s}{2})\Gamma (\frac{1-s}{2})}.
\end{equation*}

The path of integration may be deformed into one running clockwise around $%
R_{+}-c$. Comparison with the definition of the Fox $H$-function (see,
Eqs.(58) and (59), in \cite{Laskin5})) leads to

\begin{equation}
K^{(0)}(x,\tau )=  \label{eq53}
\end{equation}

\begin{equation*}
\frac 1\alpha \left( \frac \hbar {(\hbar /iD_\alpha \tau )^{1/\alpha
}}\right) ^{-1}H_{2,2}^{1,1}\left[ \frac 1\hbar \left( \frac \hbar
{iD_\alpha \tau }\right) ^{1/\alpha }x\mid \QATOP{(1-1/\alpha ,1/\alpha
),(1/2,1/2)}{(0,1),(1/2,1/2)}\right] .
\end{equation*}

Applying the Property 12.2.5, Ref.\cite{Laskin5}, of the Fox $H$-function we
can express $K^{(0)}(x,\tau )$ as

\begin{equation}
K^{(0)}(x,\tau )=\frac{1}{\alpha x}H_{2,2}^{1,1}\left[ \frac{1}{\hbar }%
\left( \frac{\hbar }{iD_{\alpha }\tau }\right) ^{1/\alpha }x\mid \QATOP{%
(1,1/\alpha ),(1,1/2)}{(1,1),(1,1/2)}\right] ,\qquad x>0.  \label{eq54}
\end{equation}

Or for any $x$,

\begin{equation}
K^{(0)}(x,\tau )=\frac{1}{\alpha x}H_{2,2}^{1,1}\left[ \frac{1}{\hbar }%
\left( \frac{\hbar }{iD_{\alpha }\tau }\right) ^{1/\alpha }|x|\mid \QATOP{%
(1,1/\alpha ),(1,1/2)}{(1,1),(1,1/2)}\right] ,  \label{eq55}
\end{equation}

Substituting $x\equiv x_{b}-x_{a},$ $\tau \equiv t_{b}-t_{a},$ finally yields

\begin{equation*}
K^{(0)}(x_{b}t_{b}|x_{a}t_{a})=
\end{equation*}

\begin{equation}
\frac 1{\alpha |x_b-x_a|}H_{2,2}^{1,1}\left[ \frac 1\hbar \left( \frac \hbar
{iD_\alpha (t_b-t_a)}\right) ^{1/\alpha }|x_b-x_a|\mid \QATOP{(1,1/\alpha
),(1,1/2)}{(1,1),(1,1/2)}\right] .  \label{eq56}
\end{equation}

This is new equation for 1D free particle fractional quantum mechanical
kernel $K^{(0)}(x_{b}t_{b}|x_{a}t_{a})$.

Let us show that Eq.(\ref{eq56}) includes as a particular case at $\alpha =2$
the well-known Feynman quantum mechanical kernel, see Eq.(3-3) in \cite%
{Feynman}. Setting in Eq.(\ref{eq56}) $\alpha =2$, applying the series
expansion for the function

\begin{equation*}
H_{2,2}^{1,1}\left[ \frac{1}{\hbar }\left( \frac{\hbar }{iD_{\alpha
}(t_{b}-t_{a})}\right) ^{1/2}|x_{b}-x_{a}|\mid \QATOP{(1,1/2),(1,1/2)}{%
(1,1),(1,1/2)}\right] ,
\end{equation*}

and finally, substituting $k\rightarrow 2l$ yield

\begin{equation}
K^{(0)}(x,\tau )|_{\alpha =2}=\frac{1}{2\hbar }\left( \frac{\hbar }{%
iD_{2}\tau }\right) ^{1/2}\sum\limits_{l=0}^{\infty }\left( -\frac{1}{\hbar }%
\left( \frac{\hbar }{iD_{2}\tau }\right) ^{1/2}\right) ^{2l}\frac{|x|^{2l}}{%
(2l)!}\frac{1}{\Gamma (\frac{1}{2}-l)}.  \label{eq57}
\end{equation}

Taking into account the identity $\Gamma (\frac{1}{2}+z)\Gamma (\frac{1}{2}%
-z)=\frac{\pi }{\cos \pi z},$ and applying the Gauss multiplication formula $%
\Gamma (2l)=\sqrt{\frac{2^{4l-1}}{2\pi }}\Gamma (l)\Gamma (l+\frac{1}{2})$,
we find that

\begin{equation}
(2l)!\Gamma (\frac{1}{2}-l)=\frac{\sqrt{\pi }}{(-1)^{l}}(2)^{2l}l!.
\label{eq58}
\end{equation}

With help of Eq.(\ref{eq58}) the kernel $K^{(0)}(x,\tau )|_{\alpha =2}$ can
be rewritten as

\begin{equation}
K^{(0)}(x,\tau )|_{\alpha =2}=\frac{1}{2\sqrt{\pi }\hbar }\left( \frac{\hbar 
}{iD_{2}\tau }\right) ^{1/2}\sum\limits_{l=0}^{\infty }\left( -\frac{1}{%
\hbar }\left( \frac{\hbar }{iD_{2}\tau }\right) ^{1/2}\right) ^{2l}\frac{%
(-1)^{l}|x|^{2l}}{2^{2l}l!}=  \label{eq59}
\end{equation}

\begin{equation*}
\frac 1{2\sqrt{\pi }\hbar }\left( \frac \hbar {iD_2\tau }\right) ^{1/2}\exp
\{-\frac 14\frac{|x|^2}{\hbar iD_2\tau }\}.
\end{equation*}

Since $D_{2}=1/2m$ we come to the Feynman kernel (see Eq.(3-3), \cite%
{Feynman})

\begin{equation*}
K^{(0)}(x,\tau )|_{\alpha =2}\equiv K_{F}^{(0)}(x,\tau )=\sqrt{\frac{m}{2\pi
i\hbar \tau }}\exp \{\frac{im|x|^{2}}{2\hbar \tau }\}.
\end{equation*}

Thus, it is shown how Feynman a free particle kernel can be derived from the
general equation (\ref{eq56}) for the fractional quantum mechanical kernel.

\section{Applications of fractional quantum mechanics}

\subsection{A free particle fractional Schr\"{o}dinger equation}

\subsubsection{Scaling properties of 1D fractional Schr\"{o}dinger equation
for a free particle}

To make general conclusions regarding solutions of 1D fractional Schr\"{o}%
dinger equation for a free particle, let's study the scaling of the
solutions. The scale transformations could be written as

\begin{equation*}
t=\lambda t^{\prime },\qquad x=\lambda ^\beta x^{\prime },\qquad D_\alpha
=\lambda ^\gamma D_\alpha ^{\prime },\qquad \psi (x,t)=\lambda ^\delta \psi
(x^{\prime },t^{\prime }),
\end{equation*}

where $\beta $, $\gamma $, $\delta $ are exponents of the scale
transformations which should leave invariant a free particle 1D fractional
Schr\"{o}dinger equation

\begin{equation}
i\hbar \frac{\partial \psi (x,t)}{\partial t}=-D_\alpha (\hbar \nabla
)^\alpha \psi (x,t),  \label{eq1331sf}
\end{equation}

and save the normalization condition $\int\limits_{-\infty }^{\infty
}dx|\psi (x,t)|^{2}=1.$ It reduces the number of exponents up to 2 and
brings the two-parameters scale transformation group

\begin{equation}
t=\lambda t^{\prime },\qquad x=\lambda ^\beta x^{\prime },\qquad D_\alpha
=\lambda ^{\alpha \beta -1}D_\alpha ^{\prime },\qquad \psi (\lambda ^\beta
x,\lambda t)=\lambda ^{-\beta /2}\psi (x,t),  \label{eq1332sf}
\end{equation}

where $\beta $ and $\lambda $ are arbitrary group parameters.

When the initial condition $\psi (x,t=0)$ is invariant under the scaling
group Eq.(\ref{eq1332sf}) then the solution of Eq.(\ref{eq1331sf}) remains
the group invariant. As an example of invariant initial condition one may
keep in mind $\psi (x,t=0)=\delta (x)$, which gives us the Green function of
1D fractional Schr\"{o}dinger equation.

To get the general scale invariant solutions of 1D fractional Schr\"{o}%
dinger equation we may use the renormalization group framework. As far as
the scale invariant solutions of Eq.(\ref{eq1331sf}) should satisfy the
identity Eq.(\ref{eq1332sf}) for any arbitrary parameters $\beta $ and $%
\lambda $, the solutions can depend on combination of $x$ and $t$ to provide
the independency of $\beta $ and $\lambda $. Thus, because of existence of
the relationships between scaling exponents, $\alpha \beta -\gamma -1=0$ and 
$\delta +\beta /2=0$ the solutions are

\begin{equation}
\psi (x,t)=\frac 1x\Phi (x/(D_\alpha t)^{\frac 1\alpha })=\frac 1{(D_\alpha
t)^{\frac 1\alpha }}\Psi (x/(D_\alpha t)^{\frac 1\alpha }),  \label{eq1333sf}
\end{equation}

where arbitrary functions $\Phi $ and $\Psi $ are determined by the
conditions, $\Phi (.)=\psi (1,.)$ and $\Psi (.)=\psi (.,1)$.

\subsubsection{Exact solution}

Following \cite{Laskin3}, \cite{GuoXu} let's solve 1D fractional Schr\"{o}%
dinger equation for a free particle (\ref{eq1331sf}) with some initial
condition $\psi _{0}(x)$

\begin{equation}
\psi (x,t=0)=\psi _{0}(x).  \label{eq1332e}
\end{equation}

Applying the Fourier transforms Eqs.(\ref{eq12}) and using the quantum Riesz
fractional derivative Eq.(\ref{eq11}) yield for the wave function $\varphi
(p,t)$ in the momentum representation,

\begin{equation}
i\hbar \frac{\partial \varphi (p,t)}{\partial t}=D_\alpha |p|^\alpha \varphi
(p,t),  \label{eq1332f}
\end{equation}

with the initial condition $\varphi _0(p)$ given by

\begin{equation}
\varphi _{0}(p)=\varphi (p,t=0)=\int\limits_{-\infty }^{\infty }dxe^{-i\frac{%
px}{\hbar }}\psi _{0}(x).  \label{eq1333f}
\end{equation}

The solution of the problem Eqs.(\ref{eq1332f}) and (\ref{eq1333f}) is

\begin{equation}
\varphi (p,t)=\exp \{-i\frac{D_\alpha |p|^\alpha t}\hbar \}\varphi _0(p),
\label{eq1332g}
\end{equation}

Therefore, the solution of 1D fractional Schr\"{o}dinger equation Eq.(\ref%
{eq1331sf}) with initial condition given by Eq.(\ref{eq1332e}) can be
presented as

\begin{equation}
\psi (x,t)=\frac{1}{2\pi \hbar }\int\limits_{-\infty }^{\infty }dx^{\prime
}\int\limits_{-\infty }^{\infty }dp\exp \{i\frac{p(x-x^{\prime })}{\hbar }-i%
\frac{D_{\alpha }|p|^{\alpha }t}{\hbar }\}\psi _{0}(x^{\prime }),
\label{eq1333h}
\end{equation}

or

\begin{equation}
\psi (x,t)=  \label{eq1333h1}
\end{equation}

\begin{equation*}
\int\limits_{-\infty }^{\infty }dx^{\prime }\frac{1}{\alpha (x-x^{\prime })}%
H_{2,2}^{1,1}\left[ \frac{1}{\hbar }\left( \frac{\hbar }{iD_{\alpha }t}%
\right) ^{1/\alpha }|x-x^{\prime }|\mid \QATOP{(1,1/\alpha ),(1,1/2)}{%
(1,1),(1,1/2)}\right] \psi _{0}(x^{\prime }).
\end{equation*}

Here we expressed the integral over $dp$ in Eq$.$(\ref{eq1333h}) in terms of 
$H_{2,2}^{1,1}$-function. If we choose the initial condition $\psi
_{0}(x)=\delta _{0}(x)$, then Eq.(\ref{eq1333h1}) gives us quantum
mechanical kernel $K^{(0)}(x,t|0,0)$ for 1D free particle fractional Schr%
\"{o}dinger equation

\begin{equation}
K^{(0)}(x,t|0,0)=\frac{1}{\alpha x}H_{2,2}^{1,1}\left[ \frac{1}{\hbar }%
\left( \frac{\hbar }{iD_{\alpha }t}\right) ^{1/\alpha }|x|\mid \QATOP{%
(1,1/\alpha ),(1,1/2)}{(1,1),(1,1/2)}\right] ,  \label{eq1333hh}
\end{equation}

or applying the Property 12.2.5 of the Fox $H$-function (see, \cite{Laskin5}%
) we can write for $K^{(0)}(x,t|0,0)$

\begin{equation}
K^{(0)}(x,t|0,0)=  \label{eq1333k}
\end{equation}

\begin{equation*}
\frac{1}{\alpha }\left( \frac{\hbar }{(\hbar /iD_{\alpha }\tau )^{1/\alpha }}%
\right) ^{-1}H_{2,2}^{1,1}\left[ \frac{1}{\hbar }\left( \frac{\hbar }{%
iD_{\alpha }\tau }\right) ^{1/\alpha }|x|\mid \QATOP{(1-1/\alpha ,1/\alpha
),(1/2,1/2)}{(0,1),(1/2,1/2)}\right] .
\end{equation*}

It is easy to see that Eqs.(\ref{eq1333h}) and (\ref{eq1333k}) are scale
invariant solutions of 1D fractional Schr\"{o}dinger equation (see, Eq.(\ref%
{eq1331sf})) for a free particle.

\subsubsection{3D generalization}

A free particle quantum dynamics in 3D is governed by following equation
(see, Eq.(\ref{eq7}))

\begin{equation}
i\hbar \frac{\partial \psi (\mathbf{r},t)}{\partial t}=D_\alpha (-\hbar
^2\Delta )^{\alpha /2}\psi (\mathbf{r},t),\qquad \psi (\mathbf{r},t=0)=\psi
_0(\mathbf{r}).  \label{eq134a}
\end{equation}

Using 3D Fourier transformed defined by Eq.(\ref{eq9}) and the definition of
3D quantum fractional Riesz derivative given by Eq.(\ref{eq8}) yield for the
wave function $\varphi (\mathbf{p},t)$ in the momentum representation,

\begin{equation}
i\hbar \frac{\partial \varphi (\mathbf{p},t))}{\partial t}=D_{\alpha }|%
\mathbf{p}|^{\alpha }\varphi (\mathbf{p},t),  \label{eq1332fh}
\end{equation}

with the initial condition $\varphi _0(\mathbf{p})$ given by

\begin{equation}
\varphi _{0}(\mathbf{p})=\varphi (\mathbf{p},t=0)=\int\limits_{-\infty
}^{\infty }d^{3}re^{-i\frac{\mathbf{pr}}{\hbar }}\psi _{0}(\mathbf{r}).
\label{eq 1333fh}
\end{equation}

Go back to Eq.(\ref{eq134a}) we can see that the solution $\psi (\mathbf{r}%
,t)$ has a form

\begin{equation*}
\psi (\mathbf{r},t)=\frac{1}{(2\pi \hbar )^{3}}\int\limits_{-\infty
}^{\infty }d^{3}r^{\prime }\int\limits_{-\infty }^{\infty }d^{3}p\exp \{i%
\frac{\mathbf{p}(\mathbf{r}-\mathbf{r}^{\prime })}{\hbar }-i\frac{D_{\alpha
}|\mathbf{p}|^{\alpha }t}{\hbar }\}\psi _{0}(\mathbf{r}^{\prime }).
\end{equation*}

The integral over $d^{3}p$ can be expressed in terms of $H_{3,3}^{1,2}$%
-function, see, for instance Eqs.(33) and (34) in.\cite{Laskin5}. Thus, the
solution of the problem Eq.(\ref{eq134a}) is

\begin{equation}
\psi (\mathbf{r},t)=  \label{eq134b}
\end{equation}

\begin{equation*}
-\frac{1}{2\pi \alpha }\int\limits_{-\infty }^{\infty }d^{3}r^{\prime }\frac{%
1}{|\mathbf{r}-\mathbf{r}^{\prime }|^{3}}H_{3,3}^{1,2}\left[ \frac{1}{\hbar }%
\left( \frac{\hbar }{iD_{\alpha }t}\right) ^{1/\alpha }|\mathbf{r}-\mathbf{r}%
^{\prime }|\mid \QATOP{(1,1),(1,1/\alpha ),(1,1/2)}{(1,1),(1,1/2),(2,1)}%
\right] \psi _{0}(\mathbf{r}^{\prime }).
\end{equation*}

Substituting into Eq.(\ref{eq134b}) $\psi _{0}(\mathbf{r})=\delta _{0}(%
\mathbf{r})$, gives us quantum mechanical kernel $K^{(0)}(\mathbf{r},t|0,0)$
for a free particle 3D fractional Schr\"{o}dinger equation

\begin{equation}
K^{(0)}(\mathbf{r},t|0,0)=-\frac{1}{2\pi \alpha }\frac{1}{|\mathbf{r}|^{3}}%
H_{3,3}^{1,2}\left[ \frac{1}{\hbar }\left( \frac{\hbar }{iD_{\alpha }t}%
\right) ^{1/\alpha }|\mathbf{r}|\mid \QATOP{(1,1),(1,1/\alpha ),(1,1/2)}{%
(1,1),(1,1/2),(2,1)}\right] .  \label{eq134gf}
\end{equation}

This is new equation for a free particle quantum mechanical 3D kernel. We
see that in comparison with 1D case 3D quantum kernel is expressed in the
terms of $H_{3,3}^{1,2}$ Fox $H$-function. In the case $\alpha =2$ we come
to the well-known equation for Feynman 3D quantum kernel $K_{F}^{(0)}(%
\mathbf{r},t|0,0)$,

\begin{equation}
K^{(0)}(\mathbf{r},t|0,0)|_{\alpha =2}\equiv K_{F}^{(0)}(\mathbf{r}%
,t|0,0)=\left( \frac{m}{2\pi i\hbar t}\right) ^{3/2}\exp \left\{ \frac{im%
\mathbf{r}|^{2}}{2\hbar t}\right\} .  \label{eq135gf}
\end{equation}

\subsection{The infinite potential well}

A particle in a one-dimensional well moves in a potential field $V(x)$ which
is zero for $-a\leq x\leq a$ and which is infinite elsewhere,

\begin{equation*}
V(x)=\infty ,\qquad x<-a\qquad \qquad (\mathrm{i})
\end{equation*}

\begin{equation}
V(x)=0,\quad -a\leq x\leq a\quad \quad \quad \ (\mathrm{ii})  \label{eq134}
\end{equation}

\begin{equation*}
V(x)=\infty ,\qquad x>a\qquad \qquad \ (\mathrm{iii})
\end{equation*}

It is evident \textit{a priori} that the spectrum will be discrete. We are
interested in the solutions of the fractional Schr\"{o}dinger equation (\ref%
{eq22}) that describe the stationary states with well-defined energies. Such
a stationary state with an energy $E$ is described by a wave function $\psi
(x,t)$ which can be written as $\psi (x,t)=\exp \{-i\frac{Et}{\hbar }\}\phi
(x),$where $\phi (x)$ is now time independent. In the regions (i) and (iii),
(see Eq.(\ref{eq134})) we have to substitute $\infty $ for $V(x)$ into Eq.(%
\ref{eq22}), and it is easily to see that here the fractional Schr\"{o}%
dinger equation can be satisfied only if we take $\phi (x)=0$. In the middle
region, (ii), the time-independent fractional Schr\"{o}dinger equation is

\begin{equation}
-D_\alpha (\hbar \nabla )^\alpha \phi (x)=E\phi (x).  \label{eq136}
\end{equation}

We can treat this as a fractional eigenvalue problem \cite{Laskin3}. Within
region (ii), the eigenfunctions are determined by Eq.(\ref{eq136}). Outside
of the region (ii), $x<-a$ and $x>a$, the eigenfunctions are zero. We want
the wave function $\phi (x)$ to be continuous everywhere, and this means
that we impose the boundary conditions $\phi (-a)=\phi (a)=0$ for the
solutions of the fractional differential equation (\ref{eq136}). Then the
solution of Eq.(\ref{eq136}) in the region (ii) can be written as

\begin{equation*}
\phi ^{\mathrm{even}}(x)=A\cos kx,\qquad \mathrm{or}\qquad \phi ^{\mathrm{odd%
}}(x)=A\sin kx,
\end{equation*}

where the following notation is introduced

\begin{equation}
k=\frac 1\hbar (\frac E{D_\alpha })^{1/\alpha },\qquad 1<\alpha \leq 2.
\label{eq137}
\end{equation}

The even (under the reflection $x\rightarrow -x$) solution $\phi ^{\mathrm{%
even}}(x)$ satisfies the boundary conditions if

\begin{equation*}
k=(2m+1)\frac \pi {2a},\quad \quad m=0,1,2,3,...
\end{equation*}

The odd (under the reflection $x\rightarrow -x$) solution $\phi ^{\mathrm{odd%
}}(x)$ satisfies the boundary conditions if

\begin{equation}
k=\frac{m\pi }a,\quad \quad m=1,2,3,...  \label{eq138}
\end{equation}

It is easy to check that the normalized solutions are

\begin{equation*}
\phi _m^{\mathrm{even}}(x)=\frac 1{\sqrt{a}}\cos \left\{ (m+\frac 12)\frac{%
\pi x}a\right\} ,
\end{equation*}

and

\begin{equation*}
\phi _m^{\mathrm{odd}}(x)=\frac 1{\sqrt{a}}\sin \frac{m\pi x}a.
\end{equation*}

The solutions $\phi ^{\mathrm{even}}(x)$ and $\phi ^{\mathrm{odd}}(x)$ have
the property that

\begin{equation*}
\int\limits_{-a}^{a}dx\phi _{m}^{\mathrm{even}}(x)\phi _{n}^{\mathrm{even}%
}(x)=\int\limits_{-a}^{a}dx\phi _{m}^{\mathrm{odd}}(x)\phi _{n}^{\mathrm{odd}%
}(x)=\delta _{mn},
\end{equation*}

\begin{equation*}
\int\limits_{-a}^{a}dx\phi _{m}^{\mathrm{even}}(x)\phi _{n}^{\mathrm{odd}%
}(x)=0,
\end{equation*}

where $\delta _{mn}$ is the Kronecker symbol and

The eigenvalues of the particle in a well with help of Eqs.(\ref{eq137}) and
(\ref{eq138}) are \cite{Laskin3}

\begin{equation}
E_n=D_\alpha \left( \frac{\pi \hbar }a\right) ^\alpha n^\alpha ,\qquad
\qquad n=1,2,3....,\qquad 1<\alpha \leq 2.  \label{eq139}
\end{equation}

It is obviously that in the Gaussian case ($\alpha =2)$ Eq.(\ref{eq139}) is
transformed to the standard quantum mechanical equation (for example, see
Eq.(20.7), Ref.\cite{Landau}) for the energy levels for a particle in a box.

The state of the lowest energy, the ground state, in the infinite potential
well is represented by the $\phi _m^{\mathrm{even}}(x)$ at $m=0$,

\begin{equation*}
\phi _{\mathrm{ground}}(x)\equiv \phi _0^{\mathrm{even}}(x)=\frac 1{\sqrt{a}%
}\cos \{\frac{\pi x}{2a}\},
\end{equation*}

and its energy is

\begin{equation}
E_{\mathrm{ground}}=D_{\alpha }\left( \frac{\pi \hbar }{2a}\right) ^{\alpha
}.  \label{eq140}
\end{equation}

\subsection{Fractional Bohr atom}

When $V(\mathbf{r})$ is the hydrogenlike potential energy

\begin{equation*}
V(\mathbf{r})=-\frac{Ze^2}{|\mathbf{r}|},
\end{equation*}

where $e$ is the electron charge, $Ze$ is the nuclear charge of the
hydrogenlike atom, we come to the eigenvalue problem for fractional
hydrogenlike atom,

\begin{equation*}
D_{\alpha }(-\hbar ^{2}\Delta )^{\alpha /2}\phi (\mathbf{r})-\frac{Ze^{2}}{|%
\mathbf{r|}}\phi (\mathbf{r})=E\phi (\mathbf{r}).
\end{equation*}

This eigenvalue problem had been solved at first in \cite{Laskin3}. The
total energy is $E=E_{kin}+V,$ where $E_{kin}$ is the kinetic energy $%
E_{kin}=D_{\alpha }|\mathbf{p}|^{\alpha },$ and $V$ is the potential energy $%
V=-\frac{Ze^{2}}{|\mathbf{r|}}$. It is well-known that if the potential
energy is a homogeneous function of the coordinates and the motion takes
place in a finite region of space, there exists a simple relation between
the time average values of the kinetic and potential energies, known as the 
\textit{virial theorem }(see, page 23, \cite{Landau1}). It follows from the
virial theorem that between average kinetic energy and average potential
energy of the system with Hamiltonian (\ref{eq3}) there exist the relation

\begin{equation}
\alpha \overline{E}_{kin}=-\overline{V},  \label{eq40b}
\end{equation}

where the average value $\overline{f}$ of any function of time is defined as

\begin{equation*}
\overline{f}=\underset{T\rightarrow \infty }{\lim }\frac
1T\int\limits_0^\infty dtf(t).
\end{equation*}

In order to evaluate the energy spectrum of the fractional hydrogenlike atom
let us remind the \textit{Niels Bohr postulates} \cite{Bohr},\cite{Bohr1}:

1. The electron moves in orbits restricted by the requirement that the
angular momentum be an integral multiple of $\hbar $, that is, for circular
orbits of radius $a_n$, the electron momentum is restricted by

\begin{equation}
pa_{n}=n\hbar ,\qquad (n=1,2,3,...),  \label{eq41b}
\end{equation}

and furthermore the electrons in these orbits do not radiate in spite of
their acceleration. They were said to be in stationary states.

2. Electrons can make discontinuous transitions from one allowed orbit
corresponding to $n=n_1$ to another corresponding to $n=n_2$, and the change
in energy will appear as radiation with frequency

\begin{equation}
\omega =\frac{E_{n_{2}}-E_{n_{1}}}{\hbar }.  \label{eq42b}
\end{equation}

An atom may absorb radiation by having its electrons make a transition to a
higher energy orbit.

Using the first Bohr's postulate and Eq.(\ref{eq40b}) yields $\alpha
D_{\alpha }\left( \frac{n\hbar }{a_{n}}\right) ^{\alpha }=\frac{Ze^{2}}{a_{n}%
},$ from which it follows the equation for the radius of the fractional Bohr
orbits

\begin{equation}
a_{n}=a_{0}n^{\frac{\alpha }{\alpha -1}},  \label{eq43b}
\end{equation}

here $a_{0}$ is the fractional Bohr radius (the radius of the lowest, $n=1,$
Bohr orbit) defined as,

\begin{equation}
a_{0}=\left( \frac{\alpha D_{\alpha }\hbar ^{\alpha }}{Ze^{2}}\right) ^{%
\frac{1}{\alpha -1}}.  \label{eq44b}
\end{equation}

By using Eq.(\ref{eq40b}) we find for the total average energy, $\overline{E}%
=(1-\alpha )\overline{E}_{kin}.$ Thus, for the energy levels of the
fractional hydrogen-like atom we have

\begin{equation}
E_{n}=(1-\alpha )E_{0}n^{-\frac{\alpha }{\alpha -1}},\qquad 1<\alpha \leq 2,
\label{eq45b}
\end{equation}

where $E_0$ is the binding energy of the electron in the lowest Bohr orbit,
that is, the energy required to put it in a state with $E=0$ corresponding
to $n=\infty $,

\begin{equation}
E_{0}=\left( \frac{(Ze^{2})^{\alpha }}{\alpha ^{\alpha }D_{\alpha }\hbar
^{\alpha }}\right) ^{\frac{1}{\alpha -1}}.  \label{eq46b}
\end{equation}

The energy $(\alpha -1)E_{0}$ can be considered as a generalization of the
Rydberg constant of standard quantum mechanics. It is easy to see that at $%
\alpha =2$ the energy $(\alpha -1)E_{0}$ is transformed into the well-known
expression for the Rydberg constant, $\mathrm{Ry}=me^{4}/2\hbar ^{2}$.

The frequency of the radiation $\omega $ associated with the transition,
say, for example from $m$ to $n$, $m\rightarrow n$, is, according to the
second Bohr postulate,

\begin{equation}
\omega =\frac{(1-\alpha )E_{0}}{\hbar }\cdot \left[ \frac{1}{n^{\frac{\alpha 
}{\alpha -1}}}-\frac{1}{m^{\frac{\alpha }{\alpha -1}}}\right] .
\label{eq47b}
\end{equation}

The new equations (\ref{eq43b})-(\ref{eq47b}) bring us fractional
generalization of the \textquotedblright Bohr atom\textquotedblright\
theory. In the special Gaussian case (standard quantum mechanics) Eqs.(\ref%
{eq43b})-(\ref{eq47b}) allow us to reproduce the well-known results of the
Bohr theory \cite{Bohr}, \cite{Bohr1}.

\subsection{Fractional oscillator}

\subsubsection{Quarkonium and fractional oscillator}

As an another physical application of the developed FQM we propose a new
fractional approach to study the quark-antiquark $q\overline{q}$ bound
states treated within the non-relativistic potential picture \cite{Laskin1}.
Note, that only for heavy quark systems (for example, charmonium $c\overline{%
c}$ or bottonium $b\overline{b}$) the non-relativistic approach can be
justified. The term quarkonium is used to denote any $q\overline{q}$ bound
state system \cite{Berkelman} in analogy to positronium in the $\mathrm{e}%
^{+}\mathrm{e}^{-}$ system. The non-relativistic potential approach remains
the most successful and simplest way to calculate and predict energy levels
and decay rates.

Thus, from stand point of \textquotedblright potential\textquotedblright\
view, we can assume that the confining potential energy of two quarks
localized say, at the space points $\mathbf{r}_{i}$ and $\mathbf{r}_{j}$ is
given by

\begin{equation}
V(|\mathbf{r}_{i}-\mathbf{r}_{j}|)=q_{i}q_{j}|\mathbf{r}_{i}-\mathbf{r}%
_{j}|^{\beta },  \label{eq48f}
\end{equation}

where $q_{i}$ and $q_{j}$ are the color charges of $i$ and $j$ quarks
respectively and the index $\beta >0$. Equation.(\ref{eq48f}) coincides with
the QCD requirements: (i) At short distances the quarks and gluons appear to
be weakly coupled; (ii) At large distances the effective coupling becomes
strong, resulting in the phenomena of quark confinement\footnote{%
The term quark confinement describes the observation that quarks do not
occur isolated in nature, but only in hadronic bound states as the colorless
objects such as baryons and mesons.}.

Considering the $N$-quarks statistical system yields the following equation
for the potential energy $U(\mathbf{r}_1,...,\mathbf{r}_N)$ of the system,

\begin{equation}
U(\mathbf{r}_{1},...,\mathbf{r}_{N})=\sum\limits_{1\leq i<j\leq N}q_{i}q_{j}|%
\mathbf{r}_{i}-\mathbf{r}_{j}|^{\beta }.  \label{eq49f}
\end{equation}

In order to illustrate the main idea, we consider the simplest case, when
color $q_{i}$ charge only can be $q$ or $-q$ and the colorless condition $%
\sum\limits_{i=1}^{N}q_{i}=0$ takes place. Using the general statistical
mechanics approach (see the Definition 3.2.1 and the Proposition 3.2.2, Ref. 
\cite{Ruelle}) we can conclude that only for $0<\beta \leq 2$ the many
particle system with the potential energy (\ref{eq49f}) will be
thermodynamically stable.

In order to study the problem of quarkonium it seems reasonable to consider
the non-relativistic FQM model with the fractional Hamiltonian operator $%
H_{\alpha ,\beta }$ defined as

\begin{equation}
H_{\alpha ,\beta }=D_{\alpha }(-\hbar ^{2}\Delta )^{\alpha /2}+q^{2}|\mathbf{%
r}|^{\beta },\quad 1<\alpha \leq 2,\quad 1<\beta \leq 2,  \label{eq50f}
\end{equation}

where $\mathbf{r}$ is 3D vector, $\Delta =\partial ^{2}/\partial \mathbf{r}%
^{2}$ is the Laplacian, and the operator $(-\hbar ^{2}\Delta )^{\alpha /2}$
is defined by Eq.(\ref{eq8}).

For the special case, when $\alpha =\beta $ the Hamiltonian operator (\ref%
{eq50f}) has a form

\begin{equation}
H_{\alpha }=D_{\alpha }(-\hbar ^{2}\Delta )^{\alpha /2}+q^{2}|\mathbf{r}%
|^{\alpha },\quad 1<\alpha \leq 2.  \label{eq51f}
\end{equation}

It is easy to see that the Hamiltonian $H_{\alpha }$ is the fractional
generalization of 3D harmonic oscillator Hamiltonian of standard quantum
mechanics. Follow \cite{Laskin1}, \cite{Laskin4}, we will call by the
fractional oscillators the quantum mechanical models with the Hamiltonians
given by the Eqs.(\ref{eq50f}) and (\ref{eq51f}).

It would be interesting to calculate the energies of the bound states and
the decay rates based on the FQM models with the Hamiltonians (\ref{eq50f})
and (\ref{eq51f}). These results could be compared with the experimental
statistical data of $J/\psi $ decays. The information on decay rates and
angular distribution provides an ideal testing ground for the fractional
models of $q\overline{q}$ bound states.

\subsubsection{Spectrum of 1D fractional oscillator in semiclassical
approximation}

The 1D fractional oscillator with the Hamilton function $H_{\alpha
}=D_{\alpha }|p|^{\alpha }+q^{2}|x|^{\beta }$ poses an interesting problem
for semiclassical treatment \cite{Laskin4}. We set the total energy equal to 
$E$, so that

\begin{equation}
E=D_{\alpha }|p|^{\alpha }+q^{2}|x|^{\beta },  \label{eq52q}
\end{equation}

whence $|p|=\left( \frac{1}{D_{\alpha }}(E-q^{2}|x|^{\beta })\right)
^{1/\alpha }.$ At the turning points $p=0$. Thus, classical motion is thus
possible in the range $|x|\leq (E/q^{2})^{1/\beta }$.

A routine use of the Bohr-Sommerfeld quantization rule \cite{Landau} yields

\begin{equation}
2\pi \hbar (n+\frac{1}{2})=\oint
pdx=4\int\limits_{0}^{x_{m}}pdx=4\int\limits_{0}^{x_{m}}D_{\alpha
}^{-1/\alpha }(E-q^{2}|x|^{\beta })^{1\alpha }dx,  \label{eq53q}
\end{equation}

where the notation $\oint $ means the integral over one complete period of
the classical motion, $x_{m}=(E/q^{2})^{1/\beta }$ is the turning point of
classical motion. To evaluate the integral in the right hand of Eq.(\ref%
{eq53q}) we introduce a new variable $y=x(E/q^{2})^{-1/\beta }$. Then we have

\begin{equation*}
\int\limits_0^{x_m}D_\alpha ^{-1/\alpha }(E-q^2|x|^\beta )^{1\alpha
}dx=\frac 1{D_\alpha ^{1/\alpha }q^{2/\beta }}E^{\frac 1\alpha +\frac 1\beta
}\int\limits_0^1dy(1-y^\beta )^{1/\alpha }.
\end{equation*}

The integral over $dy$ can be expressed in the terms of the $B$-function.
Indeed, substitution $z=y^{\beta }$ yields\footnote{%
The $B$-function is defined by
\par
\begin{equation}
B(u,v)=\int\limits_{0}^{1}dyy^{u-1}(1-y)^{v-1}.  \label{eq54b}
\end{equation}%
\par
{}}

\begin{equation}
\int\limits_{0}^{1}dy(1-y^{\beta })^{1/\alpha }=\frac{1}{\beta }%
\int\limits_{0}^{1}dzz^{\frac{1}{\beta }-1}(1-z)^{\frac{1}{\alpha }}=\frac{1%
}{\beta }B(\frac{1}{\beta },\frac{1}{\alpha }+1).  \label{eq54q}
\end{equation}

With help of Eq.(\ref{eq54q}) we rewrite Eq.(\ref{eq53q}) as

\begin{equation*}
2\pi \hbar (n+\frac 12)=\frac 4{D_\alpha ^{1/\alpha }q^{2/\beta }}E^{\frac
1\alpha +\frac 1\beta }\frac 1\beta B(\frac 1\beta ,\frac 1\alpha +1).
\end{equation*}

The above equation gives the values of the energy of stationary states for
1D fractional oscillator \cite{Laskin4},

\begin{equation}
E_{n}=\left( \frac{\pi \hbar \beta D_{\alpha }^{1/\alpha }q^{2/\beta }}{2B(%
\frac{1}{\beta },\frac{1}{\alpha }+1)}\right) ^{\frac{\alpha \beta }{\alpha
+\beta }}\cdot (n+\frac{1}{2})^{\frac{\alpha \beta }{\alpha +\beta }}.
\label{eq55q}
\end{equation}

This equation is generalized the well-known energy spectrum of the standard
quantum mechanical oscillator (see for example, \cite{Landau}) and is
transformed to it at $\alpha =2$, $\beta =2$.

We note that at

\begin{equation}
\frac{1}{\alpha }+\frac{1}{\beta }=1,  \label{eq56q}
\end{equation}%
Eq.(\ref{eq55q}) gives the equidistant energy spectrum. When $1<\alpha \leq
2 $ and $1<\beta \leq 2$ the condition given by Eq.(\ref{eq56q}) takes place
for $\alpha =2$ and $\beta =2$ only. It means that only standard quantum
mechanical oscillator has equidistant energy spectrum.

\section{Some solvable models of fractional quantum mechanics}

Let's outlook a few analytically solvable problems of fractional quantum
mechanics recently elaborated in \cite{DongXu}.

\subsection{Bound state in $\protect\delta $--potential well}

For one dimensional attractive $\delta $--potential well, $V(x)=-\gamma
\delta (x)$, ($\gamma >0$), where $\delta (x)$ is the Dirac delta function,
fractional Schr\"{o}dinger equation Eq.(\ref{eq22}) becomes

\begin{equation}
-D_{\alpha }(\hbar \nabla )^{\alpha }\phi (x)-\gamma \delta (x)\phi
(x)=E\phi (x),\qquad 1<\alpha \leq 2.  \label{eq67}
\end{equation}

In any one-dimensional attractive potential there will be a bound state,
that is $E<0$. In this case Dong and Xu \cite{DongXu}, found the energy and
the wave function of the bound state. The bound energy has a form

\begin{equation}
E=-\left( \frac{\gamma B(1/\alpha ,1-1/\alpha )}{\pi \hbar \alpha D_{\alpha
}^{1/\alpha }}\right) ^{\alpha /(\alpha -1)},\qquad 1<\alpha \leq 2,
\label{eq67d}
\end{equation}

here $B(1/\alpha ,1-1/\alpha )$ is the $B$-function defined by Eq.(\ref%
{eq54b}).

The wave function $\phi (x)$ of the bound state is

\begin{equation}
\phi (x)=-\frac{\gamma C}{2\pi \hbar ^{2}E\alpha |x|}H_{2,3}^{2,1}\left[
|x|\left( -\frac{D_{\alpha }\hbar ^{\alpha }}{E}\right) ^{-1/\alpha }\mid 
\QATOP{(1-\frac{1}{\alpha },\frac{1}{\alpha }),(\frac{1}{2},\frac{1}{2})}{%
(1,1),(1-\frac{1}{\alpha },\frac{1}{\alpha }),(\frac{1}{2},\frac{1}{2})}%
\right] ,  \label{eq67d1}
\end{equation}

here the constant $C=\phi (0)$ such, that normalization condition

\begin{equation}
\dint\limits_{-\infty }^{\infty }dx|\phi (x)|^{2}=1,  \label{eq68dd}
\end{equation}

has to be satisfied.

\subsection{Linear potential field}

Considering a particle in a linear potential field (for example, see \cite%
{Landau}, page 74), the potential function $V(x)$ can be written as:

\begin{equation}
V(x)=\{\QATOP{Fx\qquad x\geq 0,(F>0)}{\infty \qquad x<0,\qquad \quad }
\label{eq68l}
\end{equation}

and fractional Schr\"{o}dinger equation Eq.(\ref{eq22}) becomes

\begin{equation}
-D_{\alpha }(\hbar \nabla )^{\alpha }\phi (x)+Fx\phi (x)=E\phi (x),\qquad
1<\alpha \leq 2,\qquad x\geq 0.  \label{eq69l}
\end{equation}

The continuity and bounded conditions of the wave function, let us conclude
that $\phi (x)=0$, \ $x<0$. Besides, $\phi (x)$ must satisfy the boundary
conditions

\begin{equation}
\QATOP{\phi (x)=0,\qquad x=0,}{\phi (x)=0,\qquad x\rightarrow \infty .}
\label{eq70l}
\end{equation}

Then, wave function $\phi _{n}(x)$ of the quantum state with energy $E_{n}$, 
$n=1,2,3,...$is \cite{DongXu}

\begin{equation}
\phi _{n}(x)=
\end{equation}

\begin{equation}
\frac{2\pi A}{\alpha +1}H_{2,2}^{1,1}\left[ (x-\frac{E_{n}}{F})\frac{1}{%
\hbar }\left( \frac{D_{\alpha }}{(\alpha +1)F\hbar }\right) ^{-\frac{1}{%
\alpha +1}}\mid \QATOP{(1-\frac{1}{\alpha +1},\frac{1}{\alpha +1}),(\frac{%
\alpha +2}{2(\alpha +1)},\frac{\alpha }{2(\alpha +1)})}{(0,1),(\frac{\alpha
+2}{2(\alpha +1)},\frac{\alpha }{2(\alpha +1)})}\right] ,  \label{eq71l}
\end{equation}

with the constant $A$ given by

\begin{equation}
A=\frac{1}{2\pi \hbar }\left( \frac{D_{\alpha }}{(\alpha +1)F\hbar }\right)
^{-1/(\alpha +1)},  \label{eq72l}
\end{equation}

and the energy spectra $E_{n}$

\begin{equation}
E_{n}=\lambda _{n}F\hbar \left( \frac{D_{\alpha }}{(\alpha +1)F\hbar }%
\right) ^{-1/(\alpha +1)},\qquad 1<\alpha \leq 2,\qquad n=1,2,3,...,
\label{eq73l}
\end{equation}

where $\lambda _{n}$ are the solutions of the equation \cite{DongXu}

\begin{equation}
H_{2,2}^{1,1}\left[ -\lambda _{n}\mid \QATOP{(1-\frac{1}{\alpha +1},\frac{1}{%
\alpha +1}),(\frac{\alpha +2}{2(\alpha +1)},\frac{\alpha }{2(\alpha +1)})}{%
(0,1),(\frac{\alpha +2}{2(\alpha +1)},\frac{\alpha }{2(\alpha +1)})}\right]
=0.  \label{eq74l}
\end{equation}

When $\alpha =2$ Eqs.(\ref{eq71l}) and (\ref{eq73l}) turn into well-known
equations of standard quantum mechanics \cite{Landau}, \cite{DongXu}.

Other solvable physical models of fractional quantum mechanics include 1D
Coulomb potential \cite{DongXu}, a finite square potential well, dynamics in
the field of 1D lattice, penetration through a $\delta -$potential barrier,
the Dirac comb \cite{DongXu2}, the bound state problem and penetration
through double $\delta -$potential barrier \cite{Lin}.

\section{Fractional statistical mechanics}

\subsection{Density matrix}

In order to develop the fractional statistical mechanics (FSM) let us go in
Eq.(\ref{eq42}) from imaginary time to \textquotedblright inverse
temperature\textquotedblright\ $\beta =1/k_{B}T,$ where $k_{B}$ is the
Boltzmann's constant and $T$ is the temperature, $it\rightarrow \hbar \beta $%
. Then the partition function $Z$ is expressed as a trace of the density
matrix $\rho _{L}(x,\beta |x_{0})$ \cite{Laskin1}, \cite{Laskin2}

\begin{equation*}
Z=\int dx\rho _{L}(x,\beta |x)=
\end{equation*}

\begin{equation}
\int dx\int\limits_{x(0)=x(\beta )=x}^{{}}\mathrm{D}x(\tau
)\int\limits_{{}}^{{}}\mathrm{D}p(\tau )\exp \{-\frac{1}{\hbar }%
\int\limits_{0}^{\hbar \beta }du\left\{ -ip(u)\overset{\cdot }{x}%
(u)+H_{\alpha }(p(u),x(u)\right\} ,  \label{eq190}
\end{equation}

where the fractional Hamiltonian $H_{\alpha }(p,x)$ has form (\ref{eq23pi})
and $p(u),x(u)$ may be considered as paths running along on
\textquotedblright imaginary time axis\textquotedblright , $u=it$. The
exponential expression of Eq.(\ref{eq190}) is very similar to the fractional
canonical action (\ref{eq44}). Since it governs the fractional
quantum-statistical path integrals it may be called the fractional
quantum-statistical action or fractional Euclidean action, indicated by the
superscript (e),

\begin{equation}
S_{\alpha }^{(\mathrm{e})}(p,x)=\int\limits_{0}^{\hbar \beta }du\{-ip(u)%
\overset{\cdot }{x}(u)+H_{\alpha }(p(u),x(u)\}.  \label{eq190a}
\end{equation}

The parameter $u$ is not the true time in any sense. It is just a parameter
in an expression for the density matrix (see, for instance, \cite{Feynman}).
Let us call $u$ the \textquotedblright time\textquotedblright , leaving the
quotation marks to remind us that it is not real time (although $u$ does
have the dimension of time). Likewise $x(u)$ will be called the
\textquotedblright coordinate\textquotedblright\ and $p(u)$ the
\textquotedblright momentum\textquotedblright . Then Eq.(\ref{eq190}) may be
interpreted in following way: Consider all the possible paths by which the
system can travel between the initial $x(0)$ and final $x(\beta )$
configurations in the \textquotedblright time\textquotedblright\ $\hbar
\beta .$ The fractional density matrix $\rho _{L}$ is a path integral over
all possible paths, the contribution from a particular path being the
\textquotedblright time\textquotedblright\ integral of the canonical action (%
\ref{eq190a}) (considered as the functional of the path $p(u),x(u)$ in the
phase space) divided by $\hbar $. The partition function is derived by
integrating over only those paths for which initial $x(0)$ and final $%
x(\beta )$ configurations are the same and after that we integrate over all
possible initial (or final) configurations.

\subsubsection{A free particle}

The fractional density matrix $\rho _{L}^{(0)}(x,\beta |x_{0})$ of a free
particle ($V=0$) can be written as \cite{Laskin1}, \cite{Laskin2}

\begin{equation}
\rho _L^{(0)}(x,\beta |x_0)=\frac 1{2\pi \hbar }\int\limits_{-\infty
}^\infty dp\exp \left\{ i\frac{p(x-x_0)}\hbar -\beta D_\alpha |p|^\alpha
\right\} =  \label{eq191}
\end{equation}

\begin{equation*}
=\frac 1{\alpha |x-x_0|}H_{2,2}^{1,1}\left[ \frac{|x-x_0|}{\hbar (D_\alpha
\beta )^{1/\alpha }}\mid \QATOP{(1,1/\alpha ),(1,1/2)}{(1,1),(1,1/2)}\right]
,
\end{equation*}

where $H_{2,2}^{1,1}$ is the Fox $H$-function (see, \cite{Fox}-\cite{Mathai}%
).

For 1D system of space scale $\Omega $ the trace of Eq.(\ref{eq191}) reads

\begin{equation*}
Z=\int\limits_{\Omega }dx\rho _{L}^{(0)}(x,\beta |x_{0})=\frac{\Omega }{2\pi
\hbar }\frac{1}{(\beta D_{\alpha })^{1/\alpha }}\Gamma (\frac{1}{\alpha }).
\end{equation*}

When $\alpha =2$ and $D_{2}=1/2m$ Eq.(\ref{eq191}) gives the well-known
density matrix for 1D free particle (see Eq.(10-46) of Ref. \cite{Feynman}
or Eq.(2-61) of Ref. \cite{Feynman1})

\begin{equation}
\rho ^{(0)}(x,\beta |x_0)=\left( \frac m{2\pi \hbar ^2\beta }\right)
^{1/2}\exp \left\{ -\frac m{2\hbar ^2\beta }(x-x_0)^2\right\} .
\label{eq192}
\end{equation}

The Fourier representation $\rho _L^{(0)}(p,\beta |p^{\prime })$ of the
fractional density matrix $\rho _L^{(0)}(x,\beta |x_0)$ defined by

\begin{equation*}
\rho _L^{(0)}(p,\beta |p^{\prime })=\int\limits_{-\infty }^\infty dxdx_0\rho
_L^{(0)}(x,\beta |x_0)\exp \{-\frac i\hbar (px-p^{\prime }x_0)\}
\end{equation*}

can be rewritten as

\begin{equation*}
\rho _L^{(0)}(p,\beta |p^{\prime })=2\pi \hbar \delta (p-p^{\prime })\cdot
e^{-\beta D_\alpha |p|^\alpha }.
\end{equation*}

In order to obtain a formula for the fractional partition function in the
limit of fractional classical mechanics let us study the case when $\hbar
\beta $ is small. It is easy to see that the fractional density matrix $\rho
_{L}(x,\beta |x_{0})$ can be written as

\begin{equation*}
\rho _L(x,\beta |x_0)=e^{-\beta V(x_0)}\frac 1{2\pi \hbar
}\int\limits_{-\infty }^\infty dp\exp \left\{ i\frac{p(x-x_0)}\hbar -\beta
D_\alpha |p|^\alpha \right\} .
\end{equation*}

Then the partition function $Z$ in the limit of classical mechanics becomes

\begin{equation}
Z=\int\limits_{-\infty }^{\infty }dx\rho _{L}(x,\beta |x)=\frac{\Gamma
(1/\alpha )}{2\pi \hbar (\beta D_{\alpha })^{1/\alpha }}\int\limits_{-\infty
}^{\infty }dxe^{-\beta V(x)},  \label{eq193}
\end{equation}

where $\Gamma (1/\alpha )$ is the gamma function.

The partition function $Z$ given by Eq.(\ref{eq193}) is an approximation
valid if the particles of the system cannot wander very far from their
initial positions in the \textquotedblright time\textquotedblright\ $\hbar
\beta $. The limit on the distance which the particles can wander before the
approximation breaks down can be estimated from Eq.(\ref{eq191}). We see
that if the final point differs from the initial point by as mush as

\begin{equation*}
\Delta x\simeq \hbar (\beta D_\alpha )^{1/\alpha }=\hbar \left( \frac{%
D_\alpha }{kT}\right) ^{1/\alpha }
\end{equation*}

the exponential function of Eq.(\ref{eq191}) becomes greatly reduced. From
this, we can infer that intermediate points only on paths which do not
contribute greatly to the path integral of Eq.(\ref{eq191}). Thus, we
conclude that if the potential $V(x)$ does not alter very much as $x$ moves
over this distance, then the fractional classical statistical mechanics is
valid.

\subsubsection{Motion equation for the density matrix}

The density matrix $\rho _{L}(x,\beta |x_{0})$ obeys the fractional
differential equation \cite{Laskin1}, \cite{Laskin2}

\begin{equation}
-\frac{\partial \rho _L(x,\beta |x_0)}{\partial \beta }=-D_\alpha (\hbar
\nabla _x)^\alpha \rho _L(x,\beta |x_0)+V(x)\rho _L(x,\beta |x_0)
\label{eq194}
\end{equation}

or

\begin{equation}
-\frac{\partial \rho _{L}(x,\beta |x_{0})}{\partial \beta }=H_{\alpha }\rho
_{L}(x,\beta |x_{0}),\quad \rho _{L}(x,0|x_{0})=\delta (x-x_{0}),
\label{eq195}
\end{equation}%
where the fractional Hamiltonian $H_{\alpha }$ is defined by Eq.(\ref{eq14}%
). The last equation can be considered as fractional generalization of the
Bloch equation for density matrix \cite{Bloch}.

\subsubsection{3D generalization of FSM}

The above developments can be generalized to 3D dimension. It is obviously
that a free particle density matrix $\rho _{L}^{(0)}(\mathbf{r},\beta |%
\mathbf{r}_{0})$ for 3D case has a form

\begin{equation}
\rho _{L}^{(0)}(\mathbf{r},\beta |\mathbf{r}_{0})=\frac{1}{(2\pi \hbar )^{3}}%
\int d^{3}p\cdot \exp \left\{ i\frac{\mathbf{p}(\mathbf{r}-\mathbf{r}_{0})}{%
\hbar }-\beta D_{\alpha }|\mathbf{p}|^{\alpha }\right\} ,  \label{eq196}
\end{equation}

where $\mathbf{r}$, $\mathbf{r}_{0}$ and $\mathbf{p}$ are 3D vectors.

To present the density matrix $\rho _{L}(\mathbf{r},\beta |\mathbf{r}_{0})$
in the terms of the Fox $H$-function we rewrite Eq.(\ref{eq196}) as

\begin{equation*}
\rho _L^{(0)}(\mathbf{r},\beta |\mathbf{r}_0)=\frac 1{2\pi ^2\hbar ^2|%
\mathbf{r}-\mathbf{r}_0|}\int\limits_0^\infty dpp\sin (\frac{p|\mathbf{r}_b-%
\mathbf{r}_a|}\hbar )\exp \left\{ -\beta D_\alpha |\mathbf{p}|^\alpha
\right\} .
\end{equation*}

With help of the identity $\rho _{L}^{(0)}(\mathbf{r},\beta |\mathbf{r}%
_{0})=-\frac{1}{2\pi }\frac{\partial }{\partial x}\rho _{L}^{(0)}(x,\beta
|0)|_{x=|\mathbf{r}-\mathbf{r}_{0}|},$where $\rho _{L}^{(0)}(x,\beta |0)$ is
1D density matrix given by Eq.(\ref{eq191}), we find

\begin{equation}
\rho _{L}^{(0)}(\mathbf{r},\beta |\mathbf{r}_{0})=-\frac{1}{2\pi \alpha }%
\frac{1}{|\mathbf{r}-\mathbf{r}_{0}|^{3}}H_{3,3}^{1,2}\left[ \frac{|\mathbf{r%
}-\mathbf{r}_{0}|}{\hbar (D_{\alpha }\beta )^{1/\alpha }}\mid \QATOP{%
(1,1),(1,1/\alpha ),(1,1/2)}{(1,1),(1,1/2),(2,1)}\right] .  \label{eq197}
\end{equation}

This is new equation for a free particle fractional density matrix in 3D
space.

The density matrix $\rho _{L}(\mathbf{r},\beta |\mathbf{r}_{0})$ obeys the
fractional differential equation

\begin{equation}
-\frac{\partial \rho _{L}(\mathbf{r},\beta |\mathbf{r}_{0})}{\partial \beta }%
=D_{\alpha }(-\hbar ^{2}\Delta )^{\alpha /2}\rho _{L}(\mathbf{r},\beta |%
\mathbf{r}_{0})+V(\mathbf{r})\rho _{L}(\mathbf{r},\beta |\mathbf{r}_{0}),
\label{eq198}
\end{equation}

or

\begin{equation}
-\frac{\partial \rho _{L}(\mathbf{r},\beta |\mathbf{r}_{0})}{\partial \beta }%
=H_{\alpha }\rho _{L}(\mathbf{r},\beta |\mathbf{r}_{0}),\qquad \rho _{L}(%
\mathbf{r},\beta =0|\mathbf{r}_{0})=\delta (\mathbf{r-r}_{0}),  \label{eq199}
\end{equation}

where 3D fractional Hamiltonian $H_{\alpha }$ is defined by Eq.(\ref{eq3}).

Thus, the Eqs. (\ref{eq191}), (\ref{eq194})-(\ref{eq199}) are fundamental
equations of fractional statistical mechanics.


\begin{thebibliography}{99}
\bibitem{Feynman} R. P. Feynman and A.R. Hibbs, \textit{Quantum Mechanics
and Path Integrals} (McGraw-Hill, New York, 1965).

\bibitem{Laskin1} N. Laskin, Phys.Lett. A \textbf{268}, 298 (2000).

\bibitem{Levy} P. L\'evy, \textit{Th\'eorie de l'Addition des Variables
Al\'eatoires} (Gauthier-Villars, Paris, 1937).

\bibitem{Khintchine} A. Y.Khintchine and P. L\'evy, C.R. Acad. Sci. (Paris) 
\textbf{202}, 374 (1936).

\bibitem{Kac} M. Kac, in the \textit{Second Berkeley Symposium on
Mathematical Statistics and Probability} (University of California Press,
Berkeley, Calif., 1951).

\bibitem{Mandelbrot} B.B. Mandelbrot, \textit{The Fractal Geometry of Nature}
(W.H. Freeman, New York, 1982).

\bibitem{Feder} J. Feder, \textit{Fractals} (Plenum Press, New York, 1988).

\bibitem{Landau} L.D. Landau and E.M. Lifshitz, \textit{Quantum mechanics} 
\textit{(Non-relativistic Theory)}, Vol.3, 3$^{\mathrm{rd}}$ Edition, 
\textit{Course of Theoretical Physics} (Butterworth-Heinemann, Oxford, 2003).

\bibitem{Oldham} K. B. Oldham and J. Spanier, \textit{The Fractional Calculus%
}, (Academic, New York, 1974).

\bibitem{Samko} S.G. Samko, A.A. Kilbas and O.I. Marichev, \textit{%
Fractional Integrals and Derivatives, Theory and Applications}, (Gordon and
Breach, Amsterdam, 1993).

\bibitem{Zaslavsky} A.I. Saichev and G.M. Zaslavsky, Chaos \textbf{7}, 753
(1997).

\bibitem{Miller} K.S. Miller and B. Ross, \textit{An Introduction to the
Fractional Calculus and Fractional Differential Equations}, (Wiley, New
York, 1993).

\bibitem{Podlubny} I. Podlubny, \textit{Fractional Differential Equations},
(Academic, New York, 1999).

\bibitem{Laskin2} N. Laskin, Phys. Rev. \textbf{E62}, 3135 (2000).

\bibitem{Laskin3} N. Laskin, Chaos \textbf{10}, 780 (2000).

\bibitem{Laskin4} N. Laskin, Phys. Rev. \textbf{E66}, 056108 (2002).

\bibitem{Laskin5} N. Laskin, Communications in Nonlinear Science and
Numerical Simulation \textbf{12}, 2 (2007).

\bibitem{Fox} C. Fox, Trans Am Math Soc \textbf{98}, 395\textbf{\ (}1961).

\bibitem{Srivastava} H. M. Srivastava, K. C. Gupta, S.P. Goyal SP, \textit{%
The H-function of one and two variables with applications,(} New
Delhi--Madras: South Asian Publishers, 1982).

\bibitem{Mathai} A. M. Mathai and R. K. Saxena, \textit{The H-function with
Applications in Statistics and Other Disciplines}, (Wiley Eastern, New
Delhi, 1978).

\bibitem{Ruelle} D. Ruelle, \textit{Statistical Mechanics, Rigorous Results}%
, (W.A. Benjamin, Inc. New York, Amsterdam, 1969).

\bibitem{GuoXu} X. Y. Guo and M. Y. Xu, J. Math. Phys. \textbf{47}, 082104
(2006).

\bibitem{Riesz} M. Riesz, Acta Mathematica, \textbf{81},1 (1949).

\bibitem{Kleinert} H. Kleinert, \textit{Path Integrals in Quantum Mechanics,
Statistics, Polymer Physics, and Financial Markets}, (World Scientific
Publishing Co., Singapore 4$^{\mathrm{th}}$ Edition, Chapter 2, \ 2006).

\bibitem{Bohr} N. Bohr, Phil. Mag. \textbf{26}, 1, 476, 857 (1913).

\bibitem{Bohr1} N. Bohr, \textit{Collected Works}, vol.4. ed. J. Rud Nielsen
(Amsterdam, North-Holland, 1977).

\bibitem{Landau1} L. D. Landau and E. M. Lifshitz, \textit{Mechanics, }Vol.
1, 3$^{\mathrm{rd}}$ Edition, \textit{Course of Theoretical Physics }%
(Pergamon, Oxford, 1976).

\bibitem{Berkelman} K. Berkelman, Rep. Prog. Phys. \textbf{49} 1 (1986).

\bibitem{DongXu} J. Dong and M. Xu, J. Math. Phys. \textbf{48}, 072105
(2007).

\bibitem{DongXu2} J. Dong and M. Xu, J. Math. Phys. \textbf{49}, 052105
(2008).

\bibitem{Lin} A. Lin, X. Jiang, F.Miao, Journal of Shandong University
(Engineering Science), \textbf{40}, 139 (2010).

\bibitem{Feynman1} R. P. Feynman, \textit{Statistical Mechanics,} (Benjamin,
Reading, MA, 1972).

\bibitem{Bloch} F. Bloch, Zeits. f. Physick \textbf{74}, 295 (1932).
\end{thebibliography}
\end{document}